\documentclass[12pt,a4paper]{article}

\usepackage[english]{babel}

\usepackage{indentfirst}
\usepackage{amsmath}
\usepackage{amssymb}
\usepackage[pdftex]{graphicx,color}
\usepackage{slashed}
\usepackage{cite}

\usepackage{geometry}
\usepackage{ulem}

\begin{document}

\pagestyle{empty}

\begin{flushright}
SI-HEP-2020-15 

P3H-20-032

IPPP/20/27  \\[0.2cm]

\end{flushright}

\begin{center}
\textbf{\Large \boldmath 
The pion light-cone distribution amplitude  \\[1mm]
\hspace*{1.2mm} from the pion electromagnetic form factor
}
\vspace*{1.5cm}

{\large Shan Cheng$^{\rm \, a}$, Alexander Khodjamirian$^{\rm \, b}$, and Aleksey~V.~Rusov$^{\rm \, c}$}
\vspace*{0.4cm}

\textsl{%
$^{\rm \, a}$School of Physics and Electronics, Hunan University, \\
410082 Changsha, China \\[3mm]
$^{\rm \, b}$Theoretische Physik 1, Naturwissenschaftlich-Technische Fakult\"at,\\
Universit\"at Siegen, D-57068 Siegen, Germany \\[3mm]
$^{\rm \, c}$Institute for Particle Physics Phenomenology, Durham University, \\
DH1 3LE Durham, United Kingdom \\[3mm]
}
\vspace*{10mm}

\textbf{\large Abstract}\\[10pt]
\parbox[t]{0.9\textwidth}{
We suggest to probe  the pion light-cone distribution amplitude, applying a
dispersion relation for the pion electromagnetic form factor.
Instead of the standard dispersion relation, we use 
the equation between the spacelike form factor
$F_\pi(Q^2)$  and the integrated modulus of the timelike form factor.  
For $F_\pi(Q^2)$, the  QCD light-cone sum rule with a dominant twist-2
term is used. Adopting for the pion twist-2 distribution amplitude
a certain combination of the first few Gegenbauer polynomials, 
it is possible to fit their coefficients $a_{2,4,6,...}$ 
(Gegenbauer moments) from this equation, employing the measured 
pion timelike form factor.  For the exploratory fit we use 
the data of the BaBar collaboration.
The results definitely exclude the asymptotic  
twist-2 distribution amplitude.
Also the model  with a single $a_2\neq 0$   is 
disfavoured by the fit. Considering 
the models with $a_{n>2}\neq 0$, we find that the fitted values of the second
and fourth Gegenbauer moments cover the  intervals
$a_2 (1 \mbox{GeV}) = (0.22 - 0.33) $, 
$a_4 (1 \mbox{GeV}) = (0.12 - 0.25) $.
The higher  moments starting from $a_{8}$ are consistent
with zero, albeit with large uncertainties.
The spacelike pion form factor obtained in two different 
ways, from the dispersion relation and from the  
light-cone sum rule, agrees, within uncertainties, 
with the measurement by  the Jefferson Lab $F_\pi$ collaboration.
}

\vspace{2cm}
\end{center}

\newpage

\pagestyle{plain}

\section{Introduction}

The  light-cone distribution amplitudes (DAs) are the key elements in several 
QCD~methods used to describe the hard exclusive scattering and heavy hadron decays.  
For the processes involving a pion, the accuracy of this description depends first 
of all on our knowledge of the leading twist-2 pion DA.
The expansion of this DA in orthogonal Gegenbauer polynomials
reduces the necessary input to their coefficients $a_{n} ~(n=2,4,6,\dots)$ 
known as the Gegenbauer moments. 
In many practical applications only the low  moments are retained 
in the pion twist-2 DA. 
Such a parametrization is justified by the fact that the anomalous dimension 
of the multiplicatively renormalizable moment $a_n$ grows with the number $n$, 
so that the contributions of higher moments to the processes with a
large momentum scale are suppressed. 
Hence, for an accurate description of the pion DA, the  knowledge of 
the first few Gegenbauer moments at a certain normalization scale 
is of utmost importance. 

In QCD, the Gegenbauer moments are related to the vacuum-to-pion matrix elements 
of the local quark-antiquark operators with polynomial combinations of
derivatives acting on the quark field. Needless to say, these moments 
are genuinely nonperturbative objects. Currently, only the second Gegenbauer
moment $a_2$ is  accessible in the lattice QCD. 
Recently, it was obtained in Ref.~\cite{Bali:2019dqc} 
with an unprecedented accuracy (for earlier results see
Ref.~\cite{Arthur:2010xf,Braun:2015axa}). 
The moments  with $n>2$ are not accessible with these lattice computations, 
because of the growing number of derivatives in the underlying quark-antiquark operators. 
Alternative lattice QCD techniques \cite{Bali:2017gfr, Zhang:2020gaj} avoiding this problem are being developed.

The method of QCD sum rules \cite{Shifman:1978bx}
based on the two-point correlation function was pioneered 
for the calculation of Gegenbauer moments in Ref.~\cite{Chernyak:1981zz}  
and further used, e.g. in Refs.~\cite{Khodjamirian:2004ga,Ball:2006wn}. 
This method is also limited to the lowest moment $a_2$.
Extension to higher moments is possible if a nonlocal vacuum condensate 
is  introduced and modeled \cite{Mikhailov:1986be,Bakulev:2001pa}.    
A~different  approach employs the pion form factors obtained from 
the QCD light-cone sum rules (LCSRs).  
Initiated in Refs.~\cite{Balitsky:1986st,Chernyak:1990ag}, this technique enables
to calculate both the hard-scattering and soft-overlap contributions to the pion 
electromagnetic (e.m.) form factor \cite{Braun:1994ij,Braun:1999uj}
or to  the photon-pion transition form factor \cite{Khodjamirian:1997tk, Agaev:2010aq}
at large spacelike momentum transfers,
practically at $|q^2|\gtrsim 1~\mbox{GeV}^2$. 
The calculation is based on the light-cone expansion 
of a vacuum-to-pion correlation function
in terms of the pion DAs of growing twist.  
For both form factors, the dominant part of the sum rule contains the twist-2 DA 
convoluted with a calculable function.  This opens up a possibility to 
estimate or at least to constrain the Gegenbauer moments, comparing the LCSR result
for a form factor with its measurement in the spacelike region.
For the photon-pion transition form factor this analysis was done employing  
several available measurements of the $\gamma^*\gamma\to \pi^0$ process
(see e.g. Ref.~\cite{Agaev:2010aq,Mikhailov:2016klg}). 

In this paper, we concentrate on the pion e.m. form factor denoted as $F_\pi(q^2)$. 
It is possible to estimate the Gegenbauer moments, assuming a certain ansatz for 
the pion DA and  fitting the resulting LCSR for the spacelike form factor 
$F_\pi(q^2<0)$ to its measured values. 
The most accurate measurements to date have been performed
at Jefferson Lab  \cite{Huber:2008id} extracting the form factor
from the cross section of the pion electroproduction on a nucleon target. 
An earlier estimate of $a_2$ and $a_4$ fitting the LCSR to these data was 
obtained in Ref.~\cite{Khodjamirian:2011ub} 
(see also Refs.~\cite{Bijnens:2002mg, Agaev:2005gu}). 
However, the Jefferson Lab  measurements are limited to 
relatively small momentum transfers, $|q^2|\lesssim 2.5~$GeV$^2$ and depend on the model
description of the intermediate pion coupled to the 
nucleon\footnote{Note that a direct measurement of the pion spacelike e.m. form factor
via the electron-pion scattering exists \cite{Amendolia:1986wj}, 
but only at very small momentum transfers, that is, in the region where
LCSR is not applicable.}.

The pion e.m.~form factor $F_\pi(s)$  in the  timelike region $q^2 = s > 4m_\pi^2$ 
is directly accessible measuring the  $e^+e^-\to \pi^+\pi^-$ cross section at a 
given c.m. energy $\sqrt{s}$ of the $e^+e^-$ collision.
In addition to the $\rho$-resonance domain, $s\lesssim 1.0$ GeV$^2$,
which was scanned in great detail in several experiments,
quite accurate data \cite{Lees:2012cj} were obtained at larger energies, 
up to $s\simeq 9.0 $~GeV$^2$, by the BaBar collaboration, 
with the help of the initial-state radiation technique. 
One also has to mention the data of Belle collaboration
\cite{Fujikawa:2008ma} on the related (via isospin symmetry) 
pion weak vector form factor in the $\tau \to \pi \pi \nu_\tau$ decay. 

The purpose of this paper is to demonstrate that the timelike form factor  
$F_\pi(s)$   can provide an additional information about the pion DAs. 
This is achieved by combining the dispersion relation for the timelike form factor 
with the LCSR for the spacelike form factor. 
Note however that the standard dispersion relation demands the knowledge 
of the form factor imaginary part which is not directly measured
and depends on the parametrization of $F_\pi(s)$.
To avoid the uncertainty induced by the restoration of the imaginary part 
from the measured modulus of the timelike form factor,
we suggest to use a modified dispersion relation in which the spacelike form factor 
is equal to the integral over $|F_\pi(s)|^2$. 
One crucial condition for the validity of this relation is the 
(phenomenologically justified) assumption that the form factor $F_\pi(q^2)$ 
does not possess zeros in the complex plane of the variable $q^2$. 

In what follows, in Section~\ref{sect:disp} we derive and discuss 
in detail the modified dispersion relation used in our analysis. 
In Section~\ref{sec:lcsr} we present the necessary 
ingredients of the LCSR for the spacelike pion e.m. form factor. 
Section~\ref{sec:datatml}  contains the description 
of the timelike form factor data. In Section~\ref{sect:fit} 
the details of the form factor fit
to the dispersion integral and the resulting estimates of Gegenbauer 
moments are presented, 
and Section~\ref{sect:concl}  contains our conclusions.

\section{Dispersion relations for the pion form factor}
\label{sect:disp}
We use the standard definition of the pion e.m. form factor
in the spacelike region: 
\begin{equation}
\langle \pi^+(p_2)|j^{em}_\mu|\pi^+(p_1)\rangle=(p_1+p_2)_\mu F_\pi (q^2) \,,
\label{eq:Fpi}
\end{equation}
where 
$j^{em}_\mu = (1/2) \left(\bar{u}\gamma_\mu u-\bar{d}\gamma_\mu
  d\right)$ is the isovector component of the quark~e.m.~current,
and $q\!=\!p_2\!-\!p_1$ is the momentum transfer. The form factor
obeys the normalization condition $F_\pi(0)=1$, reflecting the unit electric
charge of $\pi^+$. The standard dispersion relation 
\begin{equation}
F_\pi (q^2) = \frac{1}{\pi} \int\limits_{s_0}^\infty 
ds \frac{{\rm Im} F_\pi (s)}{s - q^2- i\epsilon}\,,
\label{eq:dispFpi1}
\end{equation}
connects the spacelike pion form factor $F_\pi (q^2)$ at $q^2<0$ with the imaginary part 
of the timelike form factor $F_\pi (s)$ integrated over $s$ 
above the two-pion threshold $ s_0 \equiv 4 m_\pi^2$.
Note that subtractions are not necessary in Eq.~(\ref{eq:dispFpi1})
due to the power asymptotics of the pion form factor
predicted in QCD \cite{Chernyak:1977fk,Farrar:1979aw,Efremov:1979qk,Lepage:1980fj}:
\begin{equation}
F_\pi (q^2) \sim \frac{1}{q^2} ~~~\mbox{at}~~ |q^2|\to \infty.
\label{eq:pion-FF-asympt}
\end{equation}
In order to obtain the spacelike form factor $F_\pi (q^2<0)$ from
the dispersion relation~(\ref{eq:dispFpi1}), 
one has to extract the imaginary part of the timelike form factor 
from its modulus squared $|F_\pi(s)|^2$ measured in $e^+ e^- \to \pi^+ \pi^-$.
This extraction demands a realistic parameterization of $F_\pi(s)$, 
which satisfies the hadronic unitarity relation 
and reflects the presence of the $\rho$-resonance and its radial 
excitations in the $P$-wave channel of the pion-pion scattering.
Examples of elaborated parameterizations can be found e.g., in 
Refs.~\cite{Pich:2001pj, Hanhart:2012wi, Achasov:2011ra,Colangelo:2018mtw}.
Expressing $F_\pi(s)$ in terms of certain parameters, one has to fit them, 
comparing the resulting $|F_\pi(s)|^2$ to its measured values. 
Reconstructing $\mbox{Im}F_\pi(s)$ in this indirect way, one eventually
introduces additional uncertainties.

In this work, we suggest to use an alternative  dispersion relation  
directly expressing the spacelike form factor via the integral
over the modulus of the timelike form factor.
In the literature, this relation is called 
{\it the modulus representation} and goes back to 
Ref.~\cite{Geshkenbein:1969bb}.
To derive it, we introduce the following auxiliary function:
\begin{equation}
g_\pi (q^2) \equiv \frac{\ln F_\pi (q^2)}{q^2 \sqrt{s_0 - q^2}}.
\label{eq:g-def}
\end{equation}

This function has no singularities in the complex plane 
of the variable $q^2$, apart from the region $q^2=s>s_0$ on the real axis. 
This statement is only valid under the assumption that  the form factor $F_\pi(q^2)$
is free from zeros in the complex $q^2$ plane\,\footnote{The arguments in favour 
of this conjecture can be found in Ref.~\cite{Leutwyler:2002hm}, 
a more recent discussion of form factor zeros including relevant references can be
found in Ref.~\cite{Ananthanarayan:2011xt}.}, 
so that the logarithm in Eq.~(\ref{eq:g-def}) does not diverge at any $q^2$.
Note also that $g_\pi(0)$ is finite, 
as follows from the normalization of the pion form factor
at $q^2=0$. Furthermore, since $F_\pi(q^2)$ has a power
asymptotics,  the numerator in Eq.~(\ref{eq:g-def})
has a logarithmic behaviour. Therefore,
\begin{equation}
g_\pi(q^2) \sim \frac{1}{(q^{2})^\alpha}  ~~~\mbox{at}~~ |q^2|\to \infty\,,
\label{eq:gpi-asymp}
\end{equation}
where $\alpha>1$. This condition enables a dispersion relation for 
the function $g_\pi(q^2)$ which is derived in the same way as for the form factor.
We start from the Cauchy theorem: 
\begin{equation}
g_\pi (q^2) = \frac{1}{2 \pi \, i} \int\limits_C \! d z \frac{g_\pi(z)}{z -
  q^2}, 
\label{eq:gpi-Cauchy}
\end{equation}
where the integration contour $C$ circumvents the singularities on 
the real axis at $s>s_0$. This contour consists of
the two straight lines connected by a large circle with the radius~R 
and by a semicircle with an infinitesimal radius~$\epsilon \to 0$, 
centered at $s=s_0$. After that, we discard the integral over the circle 
at $R\to \infty$, making use of the asymptotics (\ref{eq:gpi-asymp}).
The integral over the semicircle vanishes at $\epsilon\to 0$. 
We also use the fact that the function
$g_\pi(s)$ is real valued at $s<s_0$ on the real axis 
and apply the Schwartz reflection principle:
$$
g_\pi(s + i \epsilon) - g_\pi (s- i \epsilon) = 2 i \, {\rm Im}\, 
g_\pi (s), \quad \epsilon \to 0.
$$
Finally, we are left with the relation
\begin{equation}
g_\pi (q^2) = \frac{1}{\pi} \int\limits_{s_0}^\infty 
d s \frac{{\rm Im} \, g_\pi (s)}{s - q^2- i \epsilon}.
\label{eq:dispgpi}
\end{equation}
On the real axis, at $s > s_0$, the singularities of the function $g_\pi(s)$ 
are determined by an overlap of the ``dynamical'' poles and branch points 
contributing to ${\rm Im} F_\pi(s)$  with the imaginary part of the 
square root function $\sqrt{s_0-s}$ which has a branch point at $s = s_0$.
Note  that, due to our standard choice of $i\epsilon$ in the
dispersion relation \footnote{As one can see below, the other branch with the $+i$ multiplying the square root 
would lead to an unphysical divergence of the pion form factor, 
$\lim_{q^2 \to - \infty}F_\pi(q^2)= \infty$.},
\begin{equation}
\sqrt{s_0 - (s+i\epsilon)} \big|_{\epsilon \to 0} = - i \sqrt{s - s_0}, \qquad s>s_0.
\label{eq:sqrt-conv}
\end{equation}
Representing 
$$F_\pi (s) = |F_{\pi} (s)| e^{i \delta_\pi (s)}$$
and using Eq.~(\ref{eq:sqrt-conv}), we obtain, at $s>s_0$,
\begin{equation}
\label{eq:imgpi}
{\rm Im} \, g_\pi(s) = 
{\rm Im} \left[\frac{\ln (|F_\pi (s)| e^{i \delta_\pi (s)})}{- i s \sqrt{s - s_0} } \right]
= \frac{1}{s \sqrt{s - s_0}}{\rm Im} \left[\frac{\ln |F_\pi (s)| + i \delta_\pi (s)}{ - i} \right]
= \frac{\ln |F_\pi (s)|}{s \sqrt{s -s_0}}.
\end{equation}
Finally, substituting this function in Eq.~(\ref{eq:dispgpi}), we arrive at 
\begin{equation}
\frac{\ln F_\pi (q^2)}{q^2 \sqrt{s_0 - q^2}} = \frac{1}{2 \pi} 
\int\limits_{s_0}^\infty \frac{d s \, 
\ln |F_\pi (s)|^2}{s\,\sqrt{s -s_0}  \, (s -q^2)}, 
\qquad q^2 <s_0, 
\label{eq:logFpirel}
\end{equation}
or, equivalently,
\begin{equation}
F_\pi (q^2) = \exp \left[ \frac{q^2 \sqrt{s_0 - q^2}}{2 \pi} 
\int\limits_{s_0}^\infty \frac{d s \, \ln |F_\pi (s)|^2}{s\,\sqrt{s - s_0}  \, (s -q^2)} \right],
\quad q^2 < s_0.
\label{eq:dispFpi2}
\end{equation} 
The above equation represents the modified dispersion relation we are aiming at. 
This formula was suggested and used in Ref.~\cite{Geshkenbein:1969bb}; 
other applications and modifications can be found e.g., in
Refs.~\cite{Leutwyler:2002hm,Ananthanarayan:2011xt,Ananthanarayan:2012tn}.
Note that Eq.~(\ref{eq:dispFpi2}) automatically ensures that $F_\pi (0) = 1$.
As demonstrated in Ref.~\cite{Geshkenbein:1969bb}, 
if, hypothetically, the form factor possesses zeros in the
complex $q^2$ plane, then, instead of the relation
(\ref{eq:dispFpi2}), $F_\pi(q^2<0)$ obeys narrow upper and lower bounds  that are 
again determined by the modulus of the
timelike form factor.
This case deserves a separate analysis which remains beyond the scope
of this paper.

\section{LCSR for the spacelike pion form factor}
\label{sec:lcsr}
The LCSR for the pion e.m. form factor is derived 
from  the correlation function:
\begin{eqnarray}
{\cal F}_{\rho\mu}(p,q) & = & i\int\! d^4 x \,e^{iqx}
\langle 0| T\{\bar{d}(0)\gamma_\rho\gamma_5u(0) \,  j_\mu^{\rm em}(x)\} |
\pi^+(p)\rangle 
\nonumber \\
& = &
 i p_\rho \, p_\mu \, {\cal F}((p-q)^2,Q^2)+\dots\,,
\label{eq:corrFpi}
\end{eqnarray}
where only the relevant Lorentz structure is shown, and the others are
indicated by the ellipsis. The invariant amplitude depends on the
two kinematical variables $(p-q)^2$ and $Q^2=-q^2$ and obeys a hadronic 
dispersion relation in the variable $(p-q)^2$ at fixed~$Q^2$:
\begin{equation}
{\cal F}((p-q)^2,Q^2)=\frac{2f_\pi F_\pi(Q^2)}{m_\pi^2-(p-q)^2} +
\int\limits_{(3m_\pi)^2}^\infty\! ds \frac{\rho^h(s,Q^2)}{s-(p-q)^2}\,,
\label{eq:lcsrdisp}
 \end{equation}
where the lowest pole corresponds to the intermediate one-pion state 
interpolated by the axial-vector current. The residue of the pion pole 
contains a product of the pion decay constant and the form factor we are interested in.
The integral over heavier states with the quantum numbers of the pion
or $a_1$ meson will be approximated using the quark-hadron duality.
Note that the lightest continuum state in this channel consists of three pions.

In  the correlation function (\ref{eq:corrFpi}), at large $Q^2$ and $|(p-q)^2|$,  
the product of the axial-vector and e.m. currents is expanded near 
the light-cone $x^2\simeq 0$ in a series of nonlocal operators. 
Their pion-to-vacuum matrix elements generate contributions of the pion DAs 
with growing twists $t=2,4,6$ 
\footnote{Note that in the chiral symmetry limit adopted here
the contributions of odd twists $t=3,5$ vanish.}.

The leading twist-2 term of the light-cone operator-product expansion (OPE) 
contains the pion twist-2 DA which is defined in a standard way:
\begin{eqnarray}
\langle 0| \bar d(0) \gamma_{\mu} \gamma_5 u(x) | \pi^+(p) \rangle 
&=& i p_{\mu} f_\pi \int\limits_0^1 d u \, e^{-iupx}
\varphi_{\pi}(u,\mu)\,,
\label{eq:pionDAdef}
\end{eqnarray}
where $\mu$ is the scale determined by the characteristic light-cone separation 
$\mu\sim 1/\sqrt{|x^2|}$.
The standard  expansion of the pion DA in Gegenbauer polynomials will be 
used in the following form: 
\begin{eqnarray}
\varphi_{\pi}(u,\mu)=6u\bar{u}\left(1+
\! \! \sum\limits_{n=2,4,...} \! \! a_{n}(\mu_0) 
L_{n}(\mu, \mu_0) \, C_{n}^{(3/2)} (u-\bar{u})\right),
\label{eq:pionDAdef1}
\end{eqnarray}
where $\bar{u}=1-u$.
The multiplicative renormalization from the default scale $\mu_0$ to the
variable scale $\mu$ is taken into account at LO by the logarithmic factor  
\begin{equation}
L_n (\mu, \mu_0) = 
\left[\frac{\alpha_s (\mu)}{\alpha_s (\mu_0)} \right]^{\gamma_{n}^{(0)}/\beta_0}
\end{equation}
with the anomalous dimensions given by 
\begin{equation}
\gamma_n^{(0)} =  4 C_F \left( \psi (n+2) + \gamma_E  - \frac{3}{4} 
- \frac{1}{2(n+1)(n+2)} \right), 
\end{equation}
where $\psi (n)$ is the digamma function and 
$\beta_0  = 11 - 2/3 n_f$ is the QCD  beta-function. 
Since at $\mu\to \infty$ all $L_n (\mu, \mu_0) \to 0$, 
the first  term in Eq.~(\ref{eq:pionDAdef})  determines the asymptotic shape of the~DA.

The correlation function (\ref{eq:corrFpi})
obtained in Ref.~\cite{Braun:1999uj} (see also Ref.~\cite{Bijnens:2002mg}) 
includes the twist-2 part consisting of the leading
order (LO) term and the next-to-leading order (NLO), $O(\alpha_s)$ 
radiative corrections. 
In addition, the subleading twist-4 term at LO and the twist-6 term 
in the factorizable approximation are taken into account.
Leaving aside many details, we present the 
LO twist-2 part of the invariant amplitude in Eq.~(\ref{eq:corrFpi}):
\begin{eqnarray}
{\cal F}^{\rm (tw2,LO)}((p-q)^2,Q^2)=2f_\pi\int\limits_0^1du
\frac{u \, \varphi_\pi(u)}{\bar{u} Q^2-u(p-q)^2}
\nonumber\\
= 
\int\limits_0^\infty\,\frac{ds}{s-(p-q)^2}
\left[2 f_\pi \frac{Q^2}{(Q^2+s)^2} \varphi_\pi(u(s))\right].
\label{eq:tw2lo}
\end{eqnarray}
The second equation in the above, obtained by transforming the integration 
variable: $u=Q^2/(s+Q^2)$, 
has a form of dispersion relation allowing
us to interpret the expression in the square brackets  as 
the OPE spectral density. According to the quark-hadron duality prescription,
the integral over $\rho^h(s,Q^2)$ in Eq.~(\ref{eq:lcsrdisp}) 
is replaced by the one in the second line of Eq.~(\ref{eq:tw2lo}), 
with the lower limit equal to the effective duality threshold $s_0^\pi$. 
Equating the two different representations of the 
correlation function, Eqs.~(\ref{eq:lcsrdisp}) and (\ref{eq:tw2lo}), 
subtracting the dual parts from both sides, 
performing the Borel transformation and transforming back  
to the variable $u$, we reproduce the LCSR
for the form factor in the LO, twist-2 approximation:
\begin{equation}
F_\pi^{\rm (tw2,LO)} (Q^2)=\int\limits_{u_0^\pi}^1 d u \, \varphi_\pi(u,\mu)
\exp\left(-\frac{\bar{u} \, Q^2}{u M^2}\right),
\label{eq:Fpitw2lo}
\end{equation}
where $u_0^\pi = Q^2/(s_0^\pi + Q^2)$. 
The complete  LCSR  for the pion e.m.~form factor obtained 
in Refs.~\cite{Braun:1999uj,Bijnens:2002mg} can be presented in a compact form:
\begin{equation}
F_\pi^{\rm (LCSR)} (Q^2) = 
F_\pi^{\rm (tw2,\, LO)} (Q^2) + 
F_\pi^{\rm (tw2,\, NLO)} (Q^2) +
F_\pi^{\rm (tw4,\, LO)} (Q^2) + 
F_\pi^{\rm (tw6,\, fact)} (Q^2) \,,
\label{eq:lcsrtot}
\end{equation}
where the second after Eq.~(\ref{eq:Fpitw2lo}) 
important contribution  is the NLO  twist-2 term.
It has a factorized form with respect to the pion DA: 
\begin{equation}
F_\pi^{\rm (tw2, \, NLO)}(Q^2)=
\frac{\alpha_s C_F}{4\pi}\bigg[\int\limits_{u_0^\pi}^1  
\!du\,\varphi_\pi(u,\mu){\cal F}_{\rm soft}(u,\mu,M^2,s_0^\pi)+
\int\limits^{u_0^\pi}_0 \!du\, 
\varphi_\pi(u,\mu){\cal F}_{\rm hard}(u,\mu,M^2,s_0^\pi)\bigg],
\label{eq:lcsrnlo}
\end{equation}
where the  analytical expressions for the functions ${\cal F}_{\rm soft}$
and ${\cal F}_{\rm hard}$ are given in Ref.~\cite{Braun:1999uj}.
Note that, the  first (second) term in the above equation can be interpreted as 
the  radiative correction to the soft-overlap part (\ref{eq:Fpitw2lo}) 
(as the  hard-scattering part) of the form factor. 

We substitute the expansion (\ref{eq:pionDAdef1}) in Eqs.~(\ref{eq:Fpitw2lo}) and 
(\ref{eq:lcsrnlo}) and represent the LCSR (\ref{eq:lcsrtot}) in a form containing 
a linear combination of Gegenbauer moments normalized at the default scale $\mu_0$:
\begin{equation}
F_\pi^{\rm (LCSR)} (Q^2) = 
F_\pi^{\rm (tw2, \, as)} (Q^2) + 
\sum_{n = 2,4,..} \! \! a_{n} (\mu_0) \,f_n(Q^2,\mu,\mu_0) +
F_\pi^{\rm (tw4,\, LO)} (Q^2) + 
F_\pi^{\rm (tw6,\, fact)} (Q^2) \,,
\label{eq:lcsrtot2}
\end{equation}
where the coefficient functions 
\begin{eqnarray}
f_{n}(Q^2,\mu, \mu_0) & = & 6 \, L_{n} (\mu, \mu_0) 
\Bigg\{ \int\limits_{u^\pi_0}^{1} \! d u \, u \, \bar{u} \, 
C_{n}^{(3/2)} (u-\bar{u}) \, e^{- \bar u Q^2/(u M^2)} \nonumber \\
& + & \frac{\alpha_s C_F}{4\pi}\bigg[\int\limits_{u_0^\pi}^1  
\! du \, u \, \bar{u} \,C_{n}^{(3/2)} (u-\bar{u}){\cal F}_{\rm soft}(u,\mu,M^2,s_0^\pi)
\nonumber\\
&+&
\int\limits^{u_0^\pi}_0 \!du\, 
u \, \bar{u} \,C_{n}^{(3/2)} (u-\bar{u}){\cal F}_{\rm hard}(u,\mu,M^2,s_0^\pi)\bigg]\Bigg\}\
\label{eq:fndef}
\end{eqnarray}
include the NLO corrections.
The dominant  part
of the LCSR (\ref {eq:lcsrtot2}), describing the twist-2 contribution
with the asymptotic DA, is reduced \cite{Braun:1999uj} to a compact expression
\begin{equation}
F_{\pi}^{\rm (tw2, \, as)}(Q^2) = 6 \int\limits_0^{s_0^{\pi}}\!ds\,e^{-s/M^2}
\frac{s\,Q^4}{(s+Q^2)^4}\left\{
 1+\frac{\alpha_s C_F}{4\pi}\left[\frac{\pi^2}{3}-6-\ln^2\frac{Q^2}{s} +
     \frac{s}{Q^2}+\frac{Q^2}{s}\right]\right\}   
\label{eq:SRas}
\end{equation} 
which is, as expected, independent of the factorization scale $\mu$.  
Importantly, apart from the  soft-overlap contribution 
decreasing as $\sim 1/Q^4$, the expression in Eq.~(\ref{eq:SRas}) also 
reproduces the asymptotic regime: 
\begin{equation}
\lim_{Q^2\to \infty} F^{\rm (tw2,as)}_\pi(Q^2)=\frac{8\pi\alpha_sf_\pi^2}{Q^2}
\label{eq:Fpiasy}
\end{equation}
which coincides  with the  perturbative QCD factorization formula
\cite{Chernyak:1977fk, Farrar:1979aw, Efremov:1979qk, Lepage:1980fj}
for the asymptotic pion DAs. 
As shown in Ref.~\cite{Braun:1999uj}, Eq.~(\ref{eq:Fpiasy}) 
follows from the $Q^2\to\infty$ limit of Eq.~(\ref{eq:SRas}), if we use
the conventional QCD sum rule \cite{Shifman:1978bx} 
for the square of the pion decay constant,
$$
f_\pi^2=\frac{1}{4\pi^2}\int\limits_0^{s_0^\pi}\!ds\,
e^{-s/M^2} +\dots\,,
$$
where the radiative corrections and condensate terms contain an extra
$\alpha_s$ and are absent in
our approximation.  
Importantly, at moderate $Q^2$ the soft-overlap contribution dominates
in the form factor (\ref{eq:lcsrtot2}). 

For the twist-4 contribution to the LCSR (\ref{eq:lcsrtot2})
we use the updated result  from Ref.~\cite{Bijnens:2002mg} 
obtained assuming the asymptotic form of all twist-4 DAs entering this contribution:
\begin{equation}
F_\pi^{\rm (tw4,\, LO)} (Q^2)=
\frac{40}{3} \delta_\pi^2 (\mu) \int\limits_{0}^{s_0^\pi} d s \, e^{-s/M^2}
\frac{Q^{8}}{(Q^2 + s)^6} 
\left(1 - \frac{9 s}{Q^2} + \frac{9 s^2}{Q^4} - \frac{s^3}{Q^6} \right),
\label{eq:tw4lcsr}
\end{equation}
where $\delta^2_\pi (\mu)$ denotes the normalization parameter of the twist-4 
pion DAs describing the vacuum-to-pion matrix element of 
the quark-antiquark-gluon operator
\begin{equation}
\label{eq:deltapi}
\langle 0 | \bar d\, \tilde G_{\nu \mu} \gamma^\nu u | \pi^+ (p) \rangle
= - i \delta_\pi^2 \, f_\pi p_\mu.
\end{equation}
We also have checked that adding the nonasymptotic terms \cite{Ball:2006wn}
to the twist-4 DAs  has a negligible numerical impact on the value
given by Eq.~(\ref{eq:tw4lcsr}).

Finally, for  the twist-6 term contribution to the form factor we use 
the factorization approximation, reduced \cite{Braun:1999uj} to the expression 
depending on the quark condensate density:
\begin{equation}
F_\pi^{\rm (tw6, \, fact)}(Q^2)=\frac{4\pi\alpha_sC_F}{3f_\pi^2 \, Q^4}
\langle \bar{q}q\rangle^{2}\,.
\label{eq:tw6}
\end{equation}

It is convenient to represent the complete LCSR (\ref{eq:lcsrtot2})
for the spacelike form factor in a more compact form:
\begin{equation}
F_\pi^{\rm (LCSR)} (Q^2) = F_\pi^{\rm (as)} (Q^2) + 
\sum_{n = 2,4,..} \! \! a_{n} (\mu_0) \,f_n(Q^2,\mu,\mu_0) \,,
\label{eq:lcsrtot3}
\end{equation}
where we use the notation
$$ 
F_\pi^{\rm (as)} (Q^2) = 
F_\pi^{\rm (tw2,as)} (Q^2) +F_\pi^{\rm (tw4,\, LO)} (Q^2) + 
F_\pi^{\rm (tw6,\, fact)} (Q^2) 
$$
and separate the part containing the asymptotic DA from the contribution
generated by nonasymptotic terms in the Gegenbauer expansion. 
Below, this expression will be used in the l.h.s of
the dispersion representation~(\ref{eq:dispFpi2}).
Eq.~(\ref{eq:lcsrtot3}) is valid in the spacelike region, 
practically at $Q^2=-q^2\gtrsim 1.0~\mbox{GeV}^2$. 
We also constrain the momentum transfer from above taking $Q^2\lesssim 10$ GeV$^2$.
The reasons are twofold.
Firstly,  as analysed in Ref.~\cite{Braun:1999uj}, the missing higher-order (NNLO)
perturbative corrections as well as the double logarithimic $\sim \ln^2 Q^2$ 
terms reminiscent of the Sudakov logarithms 
(such a term is already present in Eq.~(\ref{eq:SRas}))
may become numerically important in LCSR at very large $Q^2$.
Secondly, as we checked numerically, at $Q^2$ above the adopted
upper limit the dispersion integral in Eq.~(\ref{eq:dispFpi2}) becomes more sensitive 
to the timelike form factor above the region where it is measured (see
the next section). 

Another important limitation concerns the Borel parameter 
and is common for all QCD sum rules. The truncation of OPE 
on one hand and the validity of the duality approximation on the other hand 
demand that this parameter  lies within a certain interval,
typically in the ballpark of $M^2=1.0$ GeV$^2$. 
Varying this and other input parameters entering Eq.~(\ref{eq:lcsrtot3})
we estimate the uncertainty of the spacelike form factor obtained from LCSR.

\section{Pion form factor in the timelike region}
\label{sec:datatml}

\begin{figure}[h]\centering 
\includegraphics[scale=0.5]{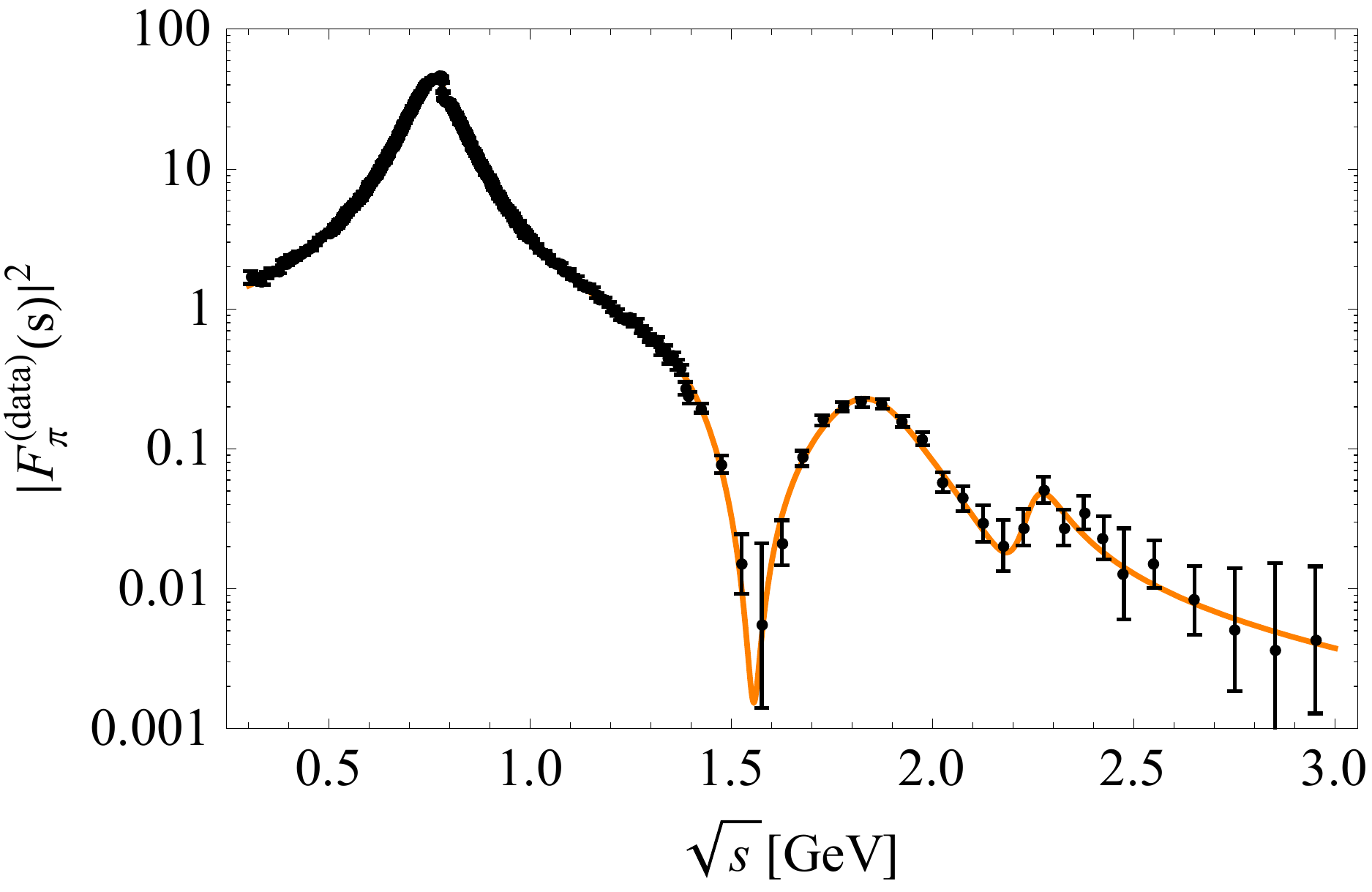}
\caption{The pion timelike form factor squared 
measured by the BaBar collaboration~\cite{Lees:2012cj}. 
The dots with error bars are the data points and the orange curve 
is the fit of the data to the resonance model~(\ref{eq:BW-model}).}
\label{fig:pion-FF-BaBar-data}
\end{figure}

Our remaining task is to specify  the modulus of the timelike pion form factor 
entering r.h.s.~of the relation (\ref{eq:dispFpi2}).
To this end we use the rich data set on the $e^+e^- \to \pi^+\pi^- (\gamma)$ 
cross section measured in the range 
$$ 0.305 \, {\rm GeV} < \sqrt{s} < 2.95 \, {\rm GeV}$$
by the BaBar collaboration \cite{Lees:2012cj}. 
We denote  by $s_{\rm max} \simeq \, 8.70~\mbox{GeV}^2$  
the upper boundary of this range and split the integration region 
in Eq.~(\ref{eq:dispFpi2}) into two intervals so that:
\begin{equation}
|F_{\pi}(s)|= \Theta(s_{\rm max} - s) \, |F^{\rm (data)}_{\pi}(s)| 
+ \Theta(s - s_{\rm max}) \, 
|F_\pi^{\rm (tail)} (s)|\,.
\label{eq:Fpi-TL-ansatz}
\end{equation}
The pion form factor extracted from the BaBar data in Ref.~\cite{Lees:2012cj} 
was fitted to a superposition of four $\rho$ resonances, 
starting from $\rho(770)$, including the three subsequent radial excitations 
and taking into account the $\rho-\omega$ mixing: 
\begin{eqnarray}
F_{\pi}^{\rm (data)}(s)  & = & 
\frac{1}{1 + 
c_{\rho'}+c_{\rho''}+c_{\rho'''}}
\Biggl[
{\rm BW}^{\rm GS}_{\rho}(s, m_\rho, \Gamma_\rho)
\frac{1 + c_\omega {\rm BW}^{\rm KS}_{\omega}(s, m_\omega, \Gamma_\omega)}
{1 + c_\omega} 
\label{eq:BW-model} \\ 
& + & c_{\rho'} {\rm BW}^{\rm GS}_{\rho'}(s, m_{\rho'}, \Gamma_{\rho'})
+c_{\rho''} {\rm BW}^{\rm GS}_{\rho''}(s, m_{\rho''}, \Gamma_{\rho''})
+c_{\rho'''} {\rm BW}^{\rm GS}_{\rho'''}(s, m_{\rho'''}, \Gamma_{\rho'''})
\Biggr].
\nonumber 
\end{eqnarray}
In this expression, all  $\rho$ resonances (the $\omega$ resonance) are described by the
Gounaris-Sakurai (GS) representation \cite{Gounaris:1968mw} 
(the K\"uhn-Santamaria (KS) representation \cite{Kuhn:1990ad}) of the Breit-Wigner function. 
In all $\rho$-resonance terms in Eq.~(\ref{eq:BW-model}),  the dependence of 
the width on the energy is taken into account and, by construction, 
the normalization condition $F_\pi (0) = 1$ is valid.
For convenience, in the Appendix we present the definitions of the BW function
used in Eq.~(\ref{eq:BW-model}). 
The numerical values of the fit parameters of resonances are taken from 
Ref.~\cite{Lees:2012cj} and shown in Table~\ref{tab:BaBar_par}.
\begin{table}[h]\centering 
\begin{tabular}{|c|c|c|c|c|}
\hline 
$R$&$m_R$ (MeV)&$\Gamma_R$(MeV)&$|c_R|$&$\phi_R$(rad)\\
\hline 
$\rho$&$775.02 \pm 0.35$&$149.59 \pm 0.67$&1.0& 0\\
\hline 
$\omega$&$781.91 \pm 0.24$&$8.13 \pm 0.45$&$(1.644 \pm 0.061) \times 10^{-3}$&$-0.011 \pm 0.037$\\
\hline 
$\rho'$&$1493 \pm 15$&$427 \pm 31$&$0.158 \pm 0.018 $&$3.76 \pm 0.10$\\
\hline 
$\rho''$&$1861 \pm 17$ &$316 \pm 26$&$0.068 \pm 0.009 $&$1.39 \pm 0.20$\\
\hline 
$\rho'''$&$2254 \pm 22$&$109 \pm 76$&$0.0051^{+0.0034}_{-0.0019} $&$0.70 \pm 0.51$\\
\hline 
\end{tabular}
\caption{Parameters of the resonances $R = \rho,\omega,\rho',\rho'',\rho'''$
in Eq.~(\ref{eq:BW-model}), obtained in Ref.~\cite{Lees:2012cj} from the fit 
to the data on the timelike pion form factor. 
The errors are added in quadrature.}
\label{tab:BaBar_par}
\end{table}
Note that the fit returns complex
valued coefficients multiplying the ${\rm BW}$ functions:
$$c_R=|c_R|e^{i\phi_R}\,, ~~~R=\omega,\rho',\rho'',\rho'''\,,$$  
implicitly reflecting a certain mixing between resonance contributions.
The timelike pion form factor measured by the BaBar collaboration 
is shown in Fig.~\ref{fig:pion-FF-BaBar-data}.

Since we are only interested in the integral over $|F_\pi(s)|$,
the expression in Eq.~(\ref{eq:BW-model}) is merely treated
as a fit function for the data points below $s_{\rm max}$. 
Moreover, this formula is not sufficiently accurate to describe
the timelike form factor at $s>s_{\rm max}$ for the  following reasons. 
Firstly, the hadronic states above the first four $\rho$ resonances 
are not taken into account, and, secondly, as we have checked 
numerically, the form factor described by Eq.~(\ref{eq:BW-model}) 
deviates from the asymptotic behaviour  $F_\pi(s)\sim 1/s$ at very large $s$. 
Hence, being continued to large negative $s=q^2\to -\infty$ it does not satisfy 
the asymptotic QCD  regime (\ref{eq:Fpiasy}), which is also obeyed by the LCSR.
Therefore, for the high-energy part of the integrand in Eq.~(\ref{eq:dispFpi2}) 
denoted as $F^{\rm (tail)}(s)$ in Eq.~(\ref{eq:Fpi-TL-ansatz}), we 
prefer to use  a theory-motivated representation of the pion form factor 
which supersedes the four-resonance formula (\ref{eq:BW-model}).

As such, we adopt the form factor suggested in Ref.~\cite{Dominguez:2001zu}
(see also Ref.~\cite{Bruch:2004py}) and inspired by the dual-resonance models 
and $N_c=\infty$ limit of QCD. The main assumption is to replace the spectrum 
of hadronic states contributing to the pion form factor 
by an infinite set of equidistant $\rho$ resonances:  
\begin{equation}
F_\pi^{(N_c=\infty)}(s)=
\sum\limits_{n=0}^{\infty} c_{n} ~ {\rm BW}_{n}(s)\,.
\label{eq:dQCD}
\end{equation}
Each resonance contribution enters the sum in a simple Breit-Wigner form 
\begin{equation}
{\rm BW}_{n}(s) = 
\frac{m_{n}^2}{m_{n}^2-s-i m_{n} \Gamma_{n}}, 
\label{eq:BW-dQCD}
\end{equation}
with a weighting coefficient:
\begin{equation}
c_{n} =  \frac{(-1)^n \Gamma(\beta-1/2)}{\alpha' m_{n}^2 \sqrt{\pi} \,
\Gamma{(n+1)}\Gamma{(\beta-1-n)}}.
\label{eq:BW-dQCD-coef} 
\end{equation}
In the above, the parameter $\alpha^\prime = 1/2 m_\rho^2$ is related 
to the $\rho$\,-meson Regge trajectory, and the masses of equidistant resonances 
are $$m_n^2 = m_\rho^2 (1 + 2 n).$$ 
In addition, in this model the total widths are assumed to grow linearly
with the resonance mass:
\begin{equation}
\Gamma_n = \gamma \, m_n \,,
\label{eq:gamn}
\end{equation}
and the parameter  $\gamma= 0.193$ is adjusted to the total width of $\rho(770)$. 
Assuming that
\begin{equation}
F^{\rm (tail)}_\pi(s) =  F_\pi^{(N_c=\infty)}(s)\,, ~~~ s\geq s_{\rm max}\,,
\label{eq:Fpitail}
\end{equation}
we fix the remaining parameter $\beta$ in Eq.~(\ref{eq:BW-dQCD-coef})
by imposing the matching condition:
\begin{equation}
|F^{\rm (data)}_\pi (s_{\rm max})| = |F^{\rm (tail)}_\pi (s_{\rm max})|\,, 
\label{eq:Fpimatch}
\end{equation}
from which we have fitted
\begin{equation}
\beta = 2.09 \pm 0.13 \,.
\label{eq:beta}
\end{equation}
In the numerical analysis, we use Eq.~(\ref{eq:dQCD})
retaining a finite sum with 100 resonance terms; 
we checked that increasing this number does not 
produce visible changes in the integral~(\ref{eq:dispFpi2}).

The advantage of 
the model~(\ref{eq:dQCD}) is that it effectively includes
all hadronic states
contributing to the form factor. In particular, the resonance widths partially account for
the continuum of intermediate hadronic states coupled to the resonances
(see Ref.~\cite{Bruch:2004py} for details). Moreover, if one introduces the correct
 threshold behavior, so that at all $\Gamma_{n}=0$ at $s<s_0=4m_\pi^2$, then 
Eq.~(\ref{eq:dQCD}) with vanishing widths is reduced to the Euler Beta
function.  
It correctly  reproduces the normalization  $F_\pi^{\rm \rm (N_c=\infty)}(0) = 1$ and
reveals the asymptotic behaviour 
$$\lim_{s\to -\infty} F_\pi^{\rm (N_c=\infty)}(s)\sim 1/s^{\beta-1},$$ 
which, having in mind the estimate (\ref{eq:beta}), is close to the QCD asymptotics.

One has to admit that the model~(\ref{eq:dQCD}) does not provide a sufficiently 
accurate description of the timelike form factor in the region below
$s_{\rm max}$ where the overlapping pattern
of the lowest $\rho$~resonances deserves a more
detailed description, and a small but visible admixture of $\omega$
has to be taken into account 
(see Table \ref{tab:BaBar_par} \cite{Lees:2012cj}). 
In~Ref.~\cite{Bruch:2004py} the $N_c=\infty$ model was modified by replacing 
the contributions of the first few resonances with GS or KS resonance formulas. 
However, we do not need such an improvement here because the
form factor $F_\pi^{(N_c=\infty)}(s)$ is used only in the region of large $s$.
\begin{figure}[t]\centering 
\includegraphics[scale=0.4]{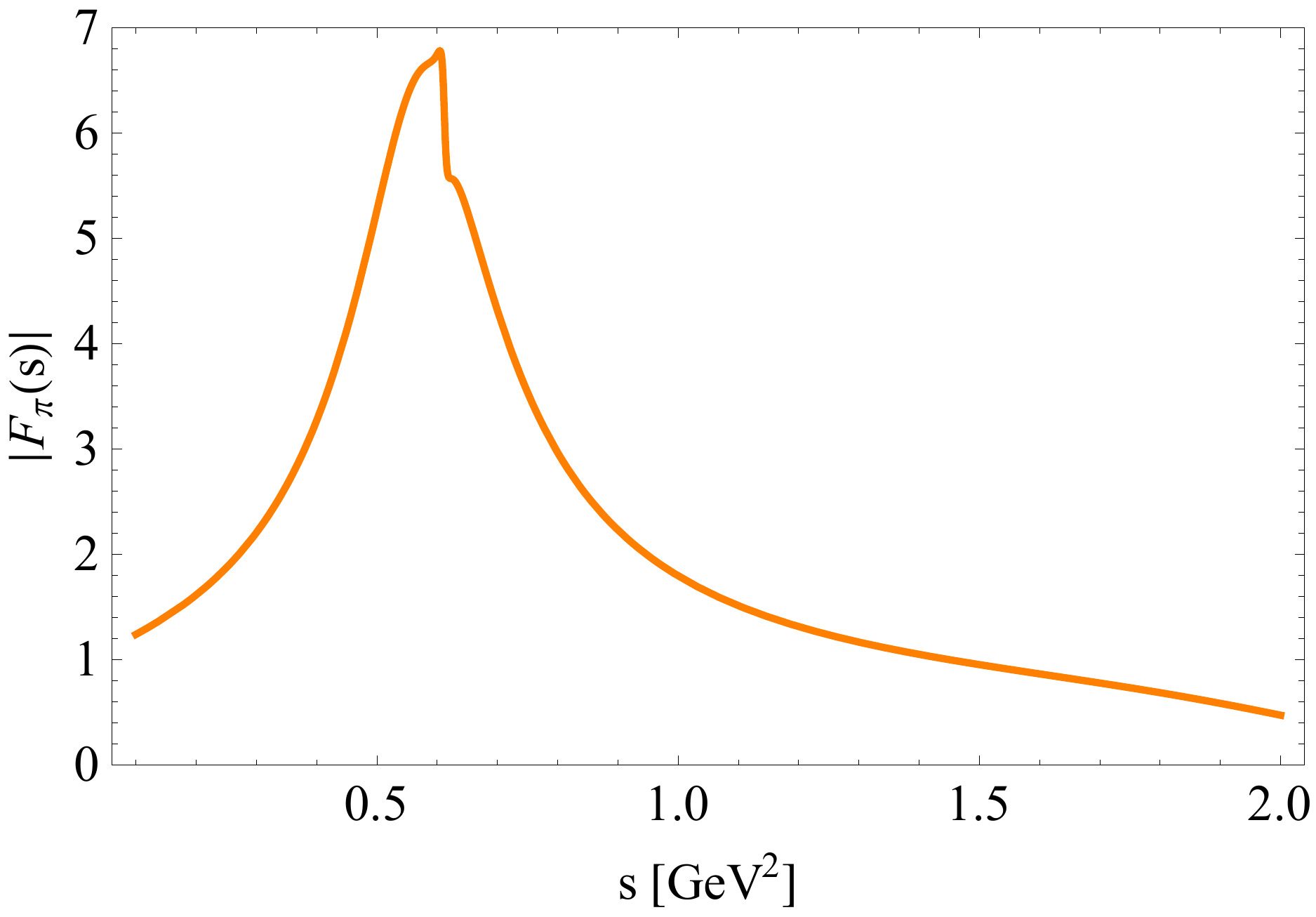} 
\includegraphics[scale=0.4]{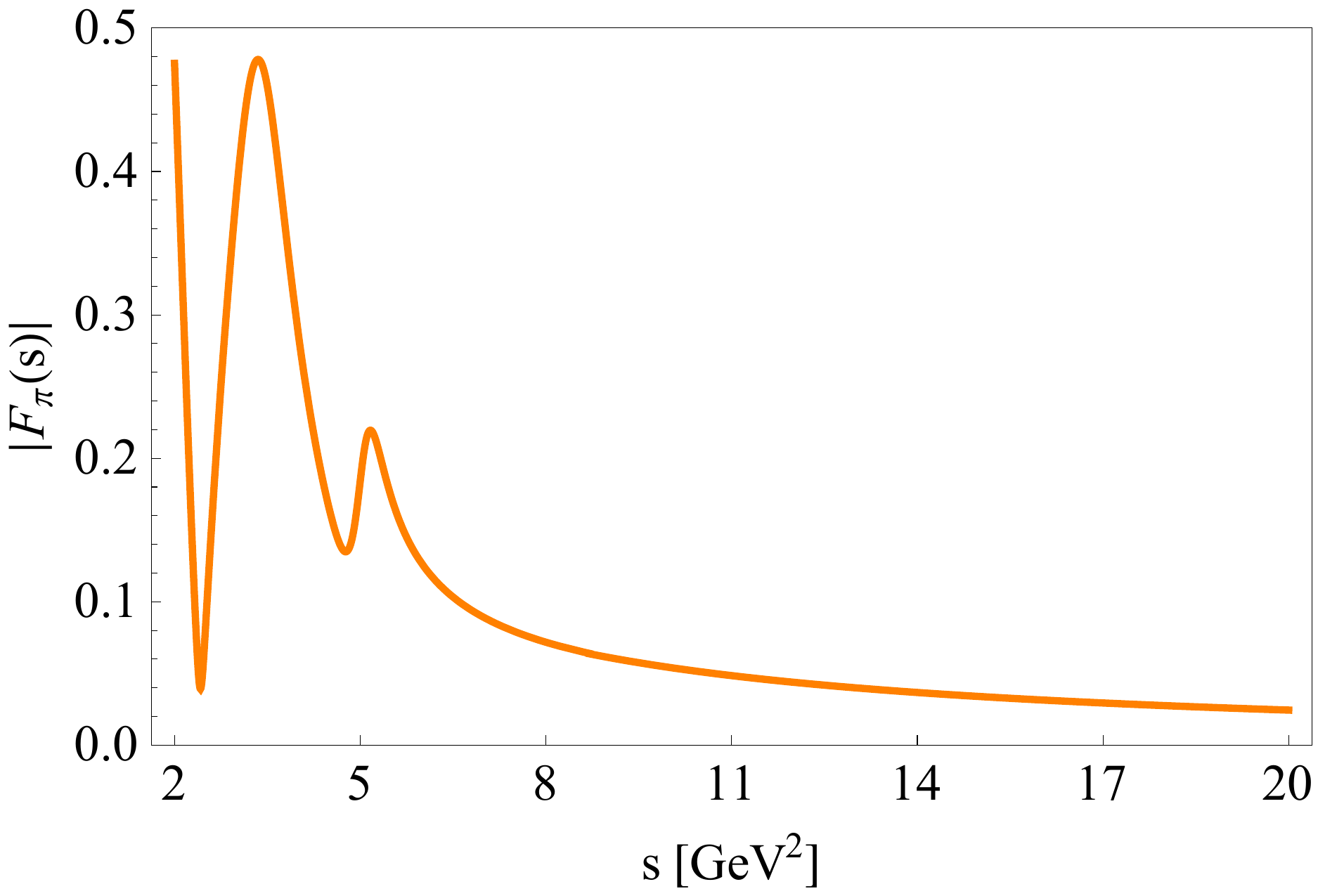} 
\caption{The modulus of the pion timelike form factor as defined in 
Eq.~(\ref{eq:Fpi-TL-ansatz}).}
\label{fig:TL-pion-FF-Abs}
\end{figure} 
In Fig.~\ref{fig:TL-pion-FF-Abs} the modulus of the timelike pion form factor 
resulting from Eq.~(\ref{eq:Fpi-TL-ansatz}) is plotted where
Eqs.~(\ref{eq:BW-model}) and (\ref{eq:Fpitail}) are used.

Our final comment concerns  the conventional  dispersion
relation (\ref{eq:dispFpi1}), which we also have probed,
numerically calculating the imaginary part of the timelike form factor from
Eq.~(\ref{eq:Fpi-TL-ansatz}).
The resulting form factor at $q^2<0$ is very close to the one obtained with 
the modulus representation. The difference varies between 
$\approx 1 \%$ for $q^2 = -1.0$ GeV$^2$ and $\approx 4 \%$
at $q^2 = -10.0$ GeV$^2$. This comparison may indicate that the zeros of  
the pion e.m. form factor  are either absent or their influence 
is beyond the accuracy of our analysis.

\section{Estimates of the Gegenbauer moments}
\label{sect:fit}

Turning to the numerical analysis, our main task is to fit 
the Gegenbauer moments from the relation (\ref{eq:dispFpi2}) rewritten as: 
\begin{equation}
F_\pi^{\rm (LCSR)} (Q^2) = \exp \left[ \frac{-Q^2\sqrt{s_0 +Q^2}}{2 \pi} 
\int\limits_{s_0}^{\infty} \frac{d s \, \ln |F_\pi (s)|^2}{s\,\sqrt{s - s_0}  \, (s +Q^2)} \right], 
\label{eq:dispFpi3}
\end{equation} 
where the l.h.s is given by Eq.~(\ref{eq:lcsrtot3}) and the integrand on r.h.s. 
by Eq.~(\ref{eq:Fpi-TL-ansatz}). 
To specify the parameters of the LCSR, 
following Ref.~\cite{Braun:1999uj}, we adopt for DAs a variable normalization scale:  
\begin{equation}
\mu^2 = (1-u) Q^2 + u M^2. 
\label{eq:scale}
\end{equation}
This choice takes into account that a typical factorization scale 
in the correlation function is determined by the interplay of 
two external variables $Q^2$ and $|(p-q)^2|$,
the latter being effectively replaced by the Borel parameter squared. 
In  Ref.~\cite{Braun:1999uj}, the two  fixed scales $\mu^2=Q^2$ 
and $\mu^2=s_0^\pi$ were also used in LCSR, 
as extreme alternatives to the default choice (\ref{eq:scale}). The 
scale dependence in the region
$Q^2\sim 1$ GeV$^2$ was found rather mild (see Fig.~4 there). Here we only use the
scale (\ref{eq:scale}), assuming that it reflects the average virtuality in the correlation function. A more
detailed analysis of scale dependence  will become possible 
if in the future the NNLO effects are taken into account.
All other input parameters in LCSR are collected in
Table~\ref{tab:parameters}. 

%
\begin{table}[h]\centering 
\begin{tabular}{|c|c|c|}
\hline 
Parameter & Value  & Reference \\
\hline 
$m_\pi$ & $139.57 $	MeV	 	     & \cite{Tanabashi:2018oca} \\
$f_\pi$ & $130.4$ MeV	 	  & \cite{Tanabashi:2018oca} \\
$s_0 ^\pi$ & $(0.7 \pm 0.1)$  GeV$^2$ & \cite{Bijnens:2002mg} \\
$M^2$     & $(1.2 \pm 0.4)$  GeV$^2$ & \cite{Bijnens:2002mg} \\
$\delta_\pi^2(1 \, {\rm GeV})$  & $(0.18 \pm 0.06)$   GeV$^2$ &\cite{Ball:2006wn}\\
$\langle \bar q q \rangle(1\,{\rm GeV}) $ &  
$- \left(269^{+15}_{-4}\right)^{3} \, $MeV$^3$
 & \cite{Tanabashi:2018oca} \\
\hline 
\end{tabular}
\caption{Numerical values of the parameters used in the LCSR 
  (\ref{eq:lcsrtot3}) for the pion spacelike form factor.}
\label{tab:parameters}
\end{table}
%
Here we follow the choice in Ref.~\cite{Bijnens:2002mg} as far as the
effective threshold and the Borel parameter interval are concerned. 
The two nonperturbative parameters, $\delta_\pi^2$ and 
$\langle \bar{q}q\rangle$, enter, respectively, 
the twist-4 and the factorizable twist-6 terms in Eq.~(\ref{eq:lcsrtot}). 
The value of $\delta_\pi^2$ determining the vacuum-to-pion 
matrix element (\ref{eq:deltapi}), is estimated from a dedicated QCD sum rule
\cite{Ball:2006wn,Novikov:1983jt} (see also Ref.~\cite{Bijnens:2002mg}). 
It agrees with the first lattice QCD determination in Ref.~\cite{Bali:2017gfr}. 
The quark condensate density is obtained from the Gell-Mann-Oakes-Renner relation: 
$\langle \bar{q}q\rangle = -f_\pi^2m_\pi^2/[2(m_u+m_d)]$ 
taking from Ref.~\cite{Tanabashi:2018oca} 
$(m_u + m_d)(2 \, {\rm GeV}) = 6.9^{+1.1}_{-0.3}\, {\rm MeV}$. 
For the quark-gluon coupling we assume the same normalization scale 
(\ref{eq:scale}) and use the average from Ref.~\cite{Tanabashi:2018oca}:
$\mathrm{\alpha_s} (M_Z)= 0.1181 \pm 0.0011$. 
For the running of $\alpha_s$ and quark masses we employ 
the RunDec code from Ref.~\cite{Chetyrkin:2000yt} so that 
e.g. $\mathrm{\alpha_s} (1 \, {\rm GeV})= 0.486 \pm 0.024 $. 
\begin{figure}[t]\centering 
\includegraphics[scale=1]{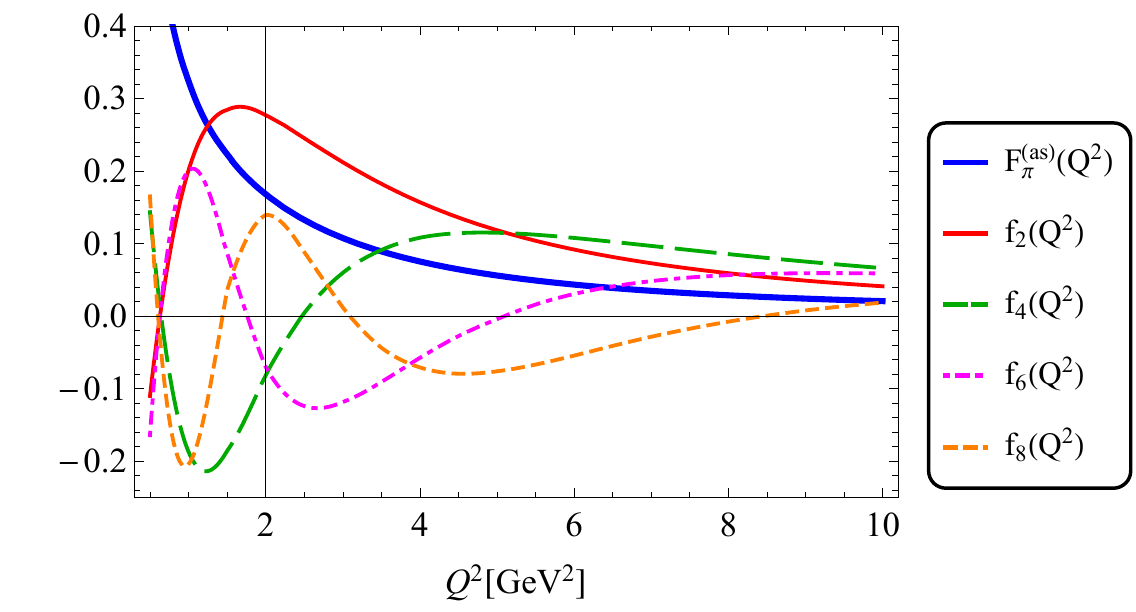}
\caption{(color online). Separate contributions to Eq.~(\ref{eq:lcsrtot3}) 
calculated at central values of the input in Table~\ref{tab:parameters} 
and at the scale (\ref{eq:scale}). 
}
\label{fig:pion-FF-LCSR}
\end{figure}

To illustrate the interplay of twist-2 contributions to 
the LCSR  (\ref{eq:lcsrtot3}), in  Fig.~\ref{fig:pion-FF-LCSR}  
we plot the form factor $F^{({\rm as})}_\pi(Q^2)$ obtained with the asymptotic DA 
including the twist-2 NLO as well as the higher-twist contributions.
Separately plotted are the coefficient functions 
$f_{n} (Q^2,\mu,\mu_0)$ at $n=2,4,6,8$.
We notice that each function $f_{ n}$ vanishes at $n$ values of
$Q^2 \ge 0$, due to the zeros of Gegenbauer polynomials. 

Having at our disposal the numerical results for both sides of Eq.~(\ref{eq:dispFpi3}) 
at multiple  values of $Q^2$ in the adopted region of validity, 
\begin{equation}
\label{eq:region}
1.0~\mbox{GeV}^2 \lesssim Q^2\lesssim 10~\mbox{GeV}^2\,,
\end{equation}
one may try to form and solve a system of linear equations for the moments $a_n$. 
However, this task it is not realistic, having in mind a limited accuracy of these equations. 
Moreover, the LCSR predictions taken at different $Q^2$ are inevitably correlated. 

To proceed, we assume a converging conformal  expansion of the DAs, so that
\begin{equation}
\label{eq:confexp}
a_{n+2}(\mu_0) <a_n(\mu_0)\,, ~~n\geq 0\,.
\end{equation}
Under this assumption, it is conceivable to adopt a certain model for 
the pion twist-2 DA with only $a_2\neq 0$ or only $a_2,a_4\neq 0$, etc. 
After that, using Eq.~(\ref{eq:dispFpi3}), it is  possible to fit 
the Gegenbauer moments within the adopted model. 
Moreover, the zeros of $f_n$ provide values of $Q^2$ at which the
contribution of the moment $a_n$ is absent.
E.g., as seen from Fig.~\ref{fig:pion-FF-LCSR},
the coefficient function $f_4$ at $Q^2\simeq 2.5 $ GeV$^2$
vanishes,  enabling one to extract $a_2$ in a model with
the two nonvanishing Gegenbauer moments $a_2,a_4\neq 0$. 
Furthermore, since the $Q^2$~dependence in
Eq.~(\ref{eq:dispFpi3}) is given in an analytical form,  
one can differentiate over $Q^2$ both parts 
of this relation yielding additional constraints. 

In this paper, we limit ourselves with an exploratory numerical study 
of Eq.~(\ref{eq:dispFpi3}) and only apply a simple fit procedure, 
assuming that the pion twist-2 DA consists of a certain combination
of few first Gegenbauer polynomials. 

\begin{figure}[t]\centering 
\includegraphics[scale=0.5]{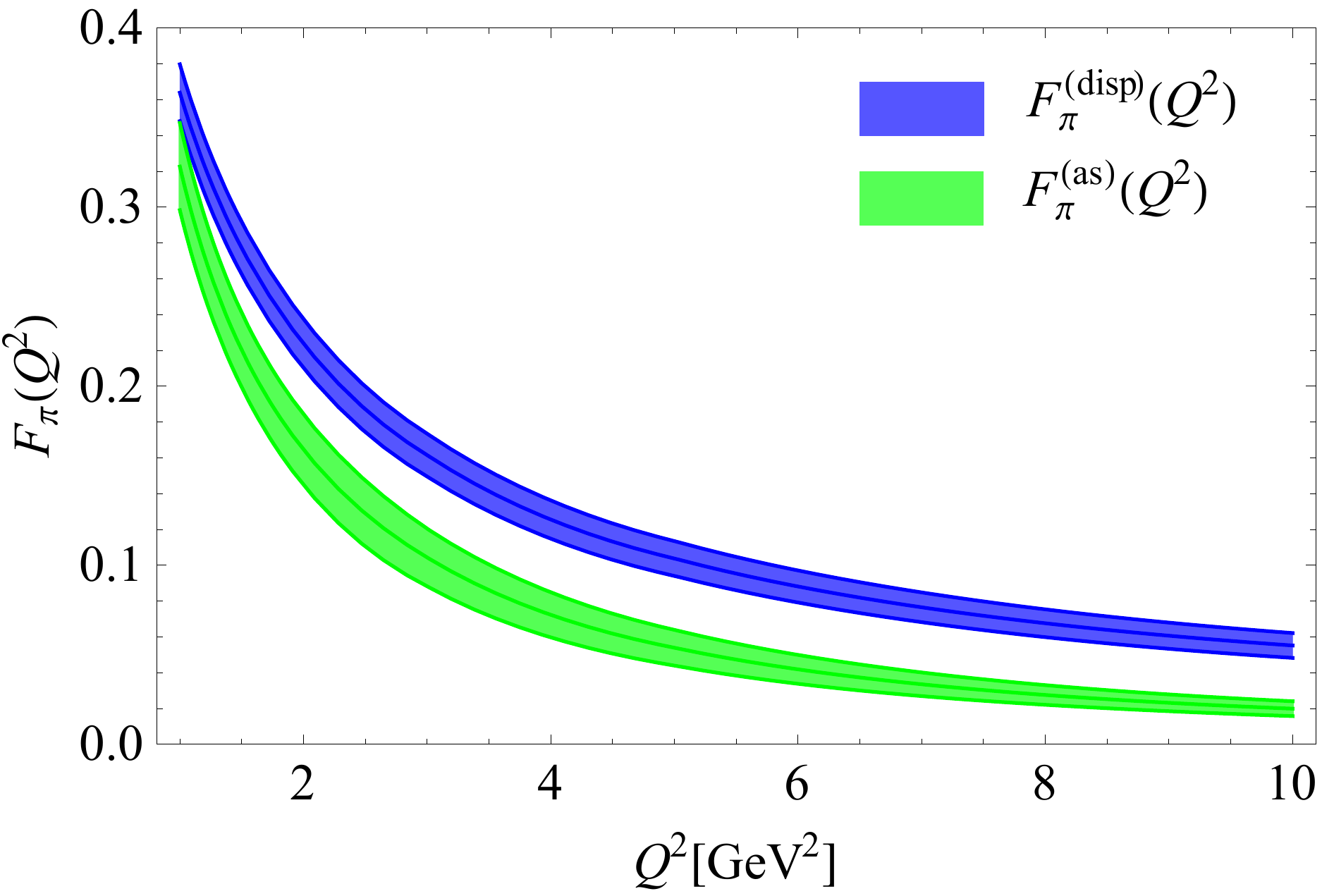}
\caption{(color online). 
The pion spacelike form factor calculated from the LCSR with the asymptotic 
twist-2 pion DA (green band) and from the dispersion relation (\ref{eq:dispFpi2}) 
(blue band) using the timelike form factor specified in Eq.~(\ref{eq:Fpi-TL-ansatz}).
}
\label{fig:LCSR-vs-BaBar}
\end{figure}

Before presenting the fit results, let us make the following observation.    
In Fig.~\ref{fig:LCSR-vs-BaBar}, the l.h.s of Eq.~(\ref{eq:dispFpi3}) 
calculated assuming the asymptotic twist-2 DA (i.e., at all $a_n=0$), 
is compared with the r.h.s. obtained from Eq.~(\ref{eq:Fpi-TL-ansatz}). 
An apparent discrepancy clearly indicates that we have to include 
nonasymptotic terms in the pion twist-2 DA. 
To this end, 
we consider four different models of this DA, retaining in the expansion
(\ref{eq:pionDAdef1}) from one to four nonvanishing Gegenbauer moments. 
For brevity we denote these models as
\begin{equation}
\{a_2\}, ~\{a_2,a_4\},~ \{a_2,a_4,a_6\},~\{a_2,a_4,a_6,a_8\}\,.
\label{eq:models}
\end{equation}
For each model we introduce the $\chi^2$-distribution function:
\begin{equation}
\chi^2 = 
\sum_{i = 1}^{N_p} 
\frac{1}{\sigma_i^2}\left[
\sum\limits_{n = 2,4,..}^{n_{\rm max}} \! \! a_{n} (\mu_0) \,f_n(Q^2_i, 
\mu_0)+F_\pi^{\rm (as)} (Q^2_i)
- F_{\pi}^{\rm (disp)} (Q^2_i)\right]^2,
\label{eq:chi-sq-function}
\end{equation}
quantifying the difference between the l.h.s. of
Eq.~(\ref{eq:dispFpi3}) (decomposed according to
Eq.~(\ref{eq:lcsrtot3})) and the r.h.s of the same
equation denoted as $F_{\pi}^{\rm (disp)}(Q^2_i)$.
In the sum in Eq.~(\ref{eq:chi-sq-function}), we include $N_p = 7$ 
"data" points at $Q^2 = \{1.0, 1.5, 2.0, 3.0, 5.0, 7.0, 9.0\} \, {\rm GeV}^2$
effectively covering the region (\ref{eq:region}).
The weighting coefficients $\sigma_i^2$ are calculated according to:
\begin{equation}
\sigma_i=\sqrt{[\Delta F_\pi^{\rm (LCSR)} (Q_i^2, a_n (\mu_0) = 0)]^2+[\Delta F_{\pi}^{\rm (disp)}
(Q^2_i)]^2}\,,
\label{eq:sigma}
\end{equation}
where the uncertainty $\Delta F_\pi^{\rm (LCSR)} (Q_i^2, a_n(\mu_0) = 0)$ attributed 
to LCSR at the point $Q_i^2$ is calculated at vanishing values of the Gegenbauer 
moments $a_n (\mu_0)$, varying all input parameters within their intervals.
The uncertainties due to the coefficient functions
$f_n$ are neglected. We have checked that their addition
to the weighting coefficients produces numerically insignificant changes 
in the fit results.
The uncertainty $\Delta F_{\pi}^{\rm (disp)}(Q_i^2)$ of the dispersion relation 
is estimated, varying the fit parameters quoted in Table~\ref{tab:BaBar_par}
within their errors (assumed uncorrelated).
We also add to this experiment-induced uncertainty the one 
due to model-dependence of the high-energy tail in $F_{\pi}^{\rm (disp)}(Q_i^2)$. 
The latter uncertainty is estimated, artificially continuing the BaBar fit 
formula for $F_\pi^{\rm(data)}(s)$ into the region $s>s_{\rm max}$ and calculating 
the variation of the dispersion integral caused by this modification. 
We have found that this variation is limited by $\pm 5 \%$.

\begin{figure}[th]\centering 
\includegraphics[scale=0.4]{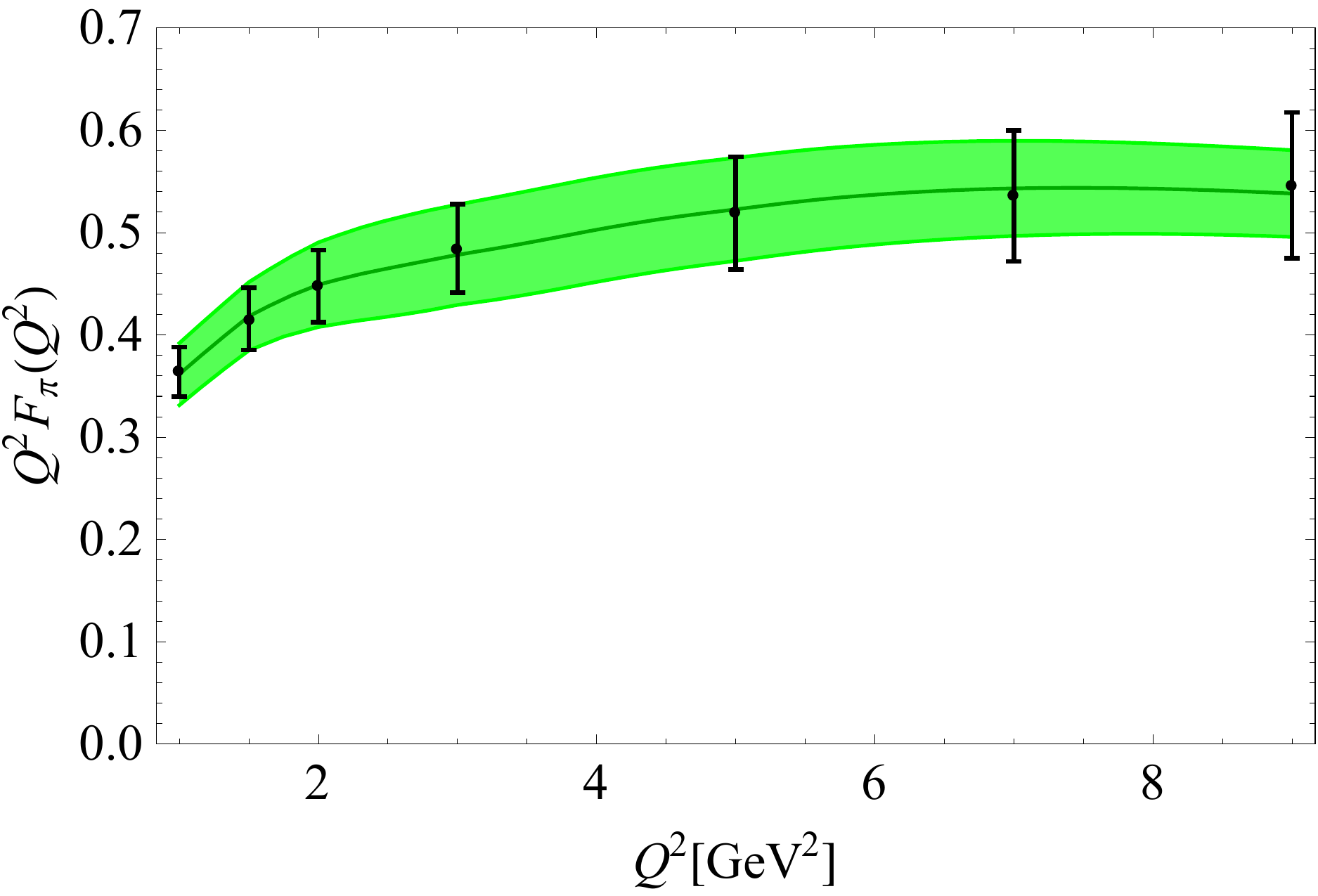}
\includegraphics[scale=0.4]{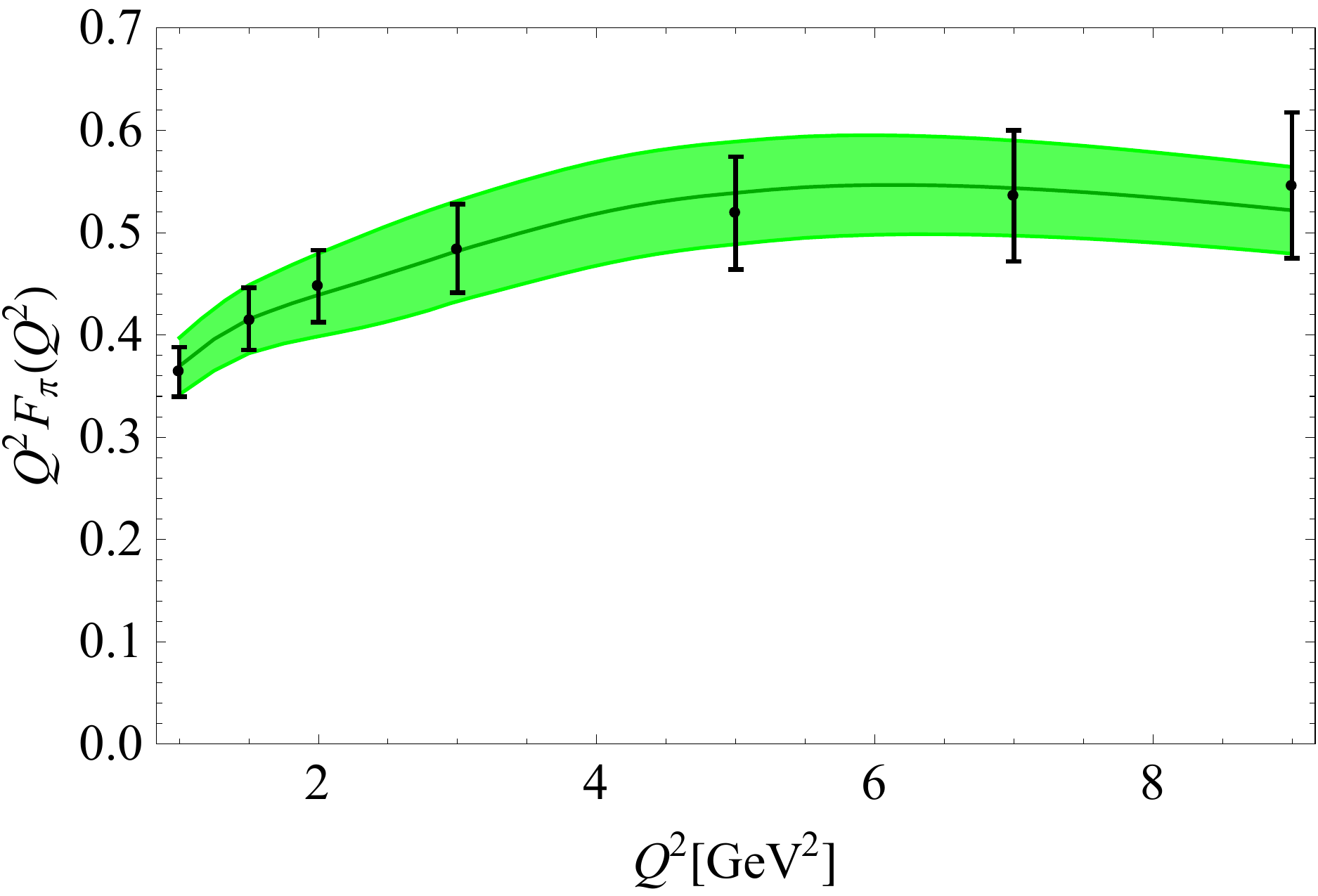}
\includegraphics[scale=0.4]{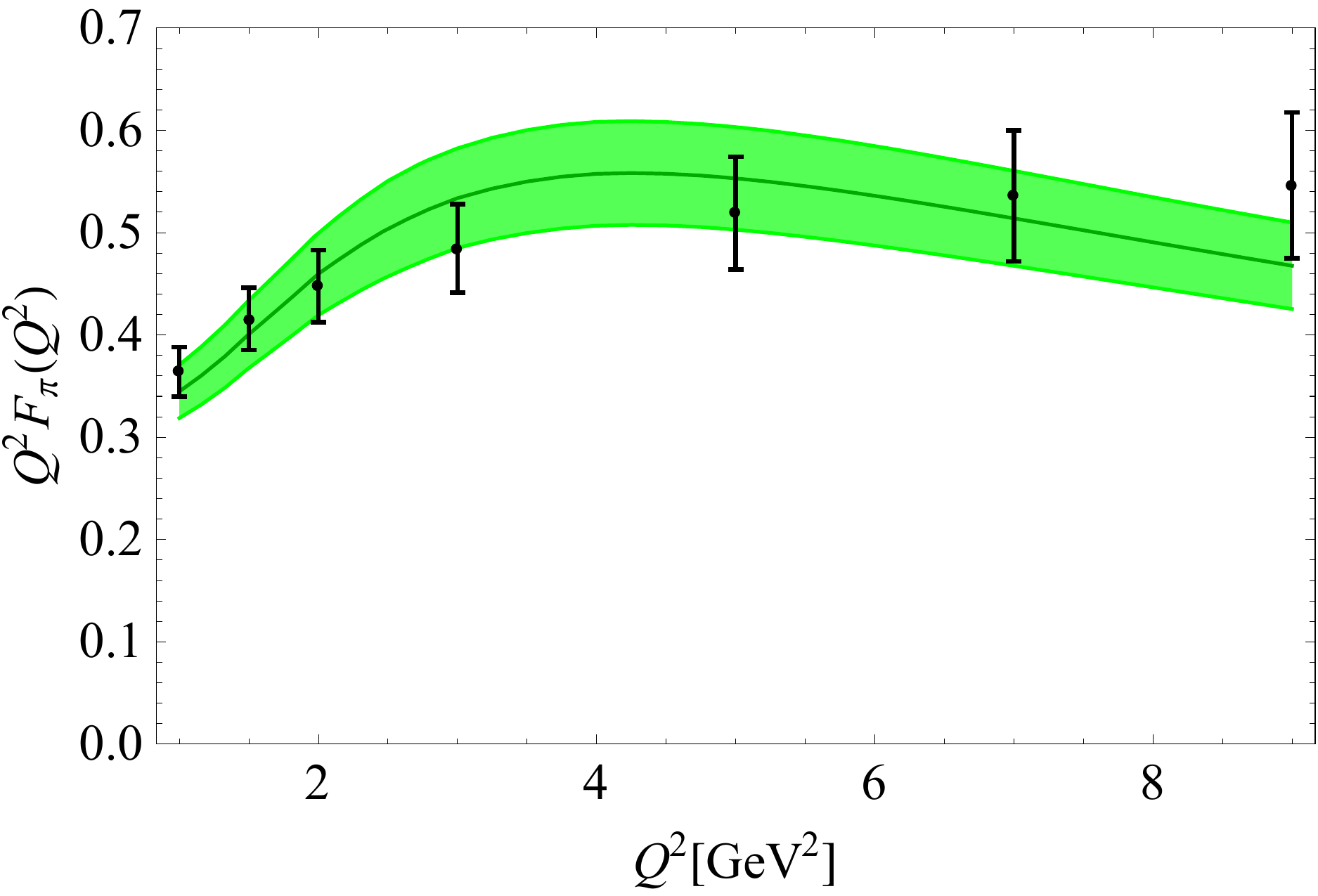}
\includegraphics[scale=0.4]{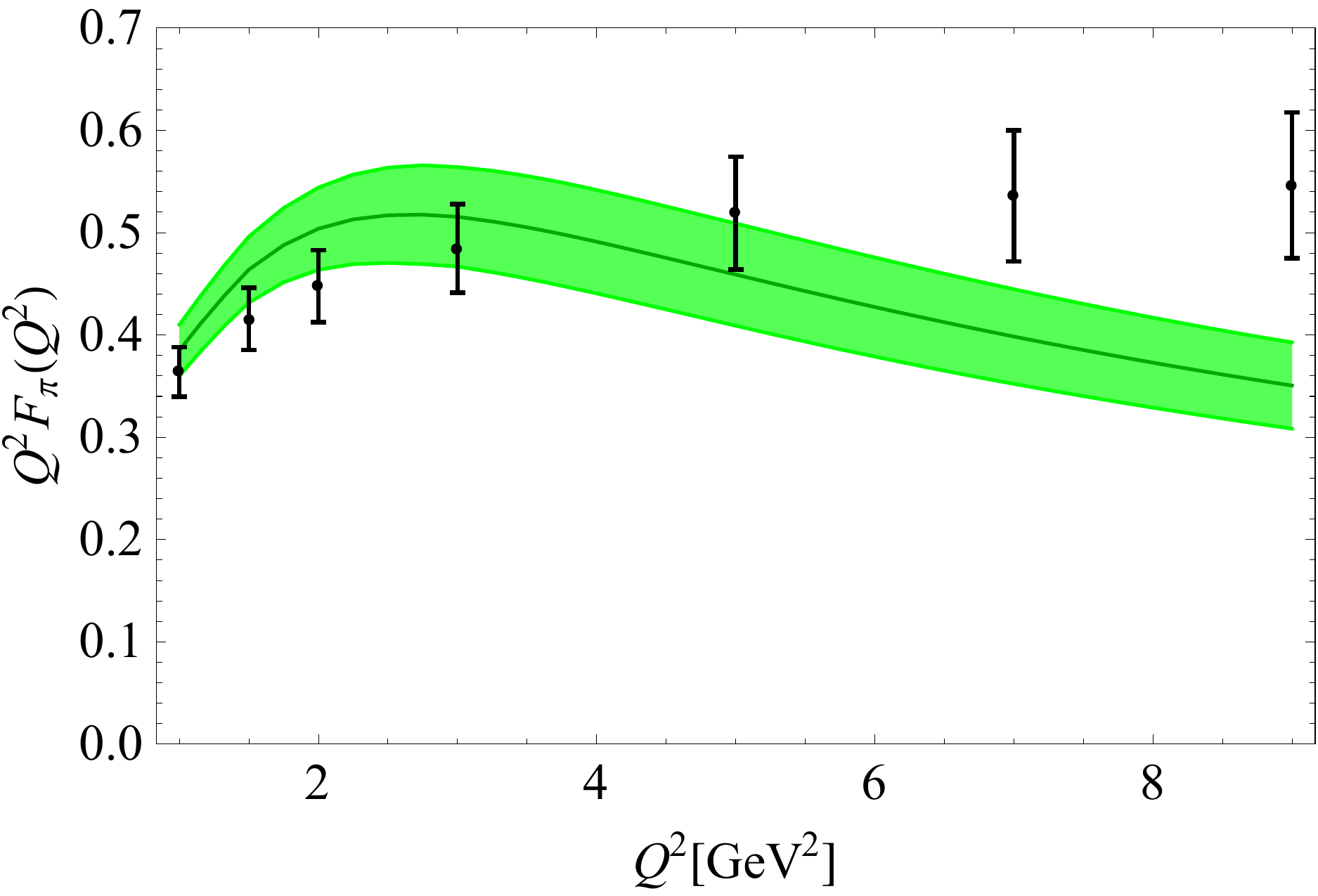}
\caption{The pion spacelike form factor calculated from the dispersion
relation (\ref{eq:dispFpi3}) (``data'' points with error bars)
and from the LCSR with the Gegenbauer
moments obtained from the fit (green bands) for the models
$\{a_2, a_4, a_6, a_8\}$, $\{a_2, a_4, a_6\}$,$\{a_2, a_4\}$ and
$\{a_2\}$, shown, respectively, on the top left, top right, bottom left and
bottom right panel.}
\label{fig:fit-result-a8-a6-a4-a2}
\end{figure}

Minimizing the $\chi^2$-distribution function~(\ref{eq:chi-sq-function}),
we obtain the estimates of the Gegenbauer moments $a_n(\mu_0)$ for all models 
listed in Eq.~(\ref{eq:models}). 
Our main numerical results are presented in Table~\ref{tab:an-res} and 
the corresponding plots are shown in Fig.~\ref{fig:fit-result-a8-a6-a4-a2}. 
\begin{table}[th]
\centering 
{\small 
\begin{tabular}{|c| c| c| c| c|c|} 
\hline 
Model &
$a_2 (1 \, {\rm GeV})$ & 
$a_4 (1 \, {\rm GeV})$ & 
$a_6 (1 \, {\rm GeV})$ & 
$a_8 (1 \, {\rm GeV})$ & 
$\mathrm{\chi^2_{\rm min}/ndf}$ 
\\
\hline 
$\{a_2\}$ & 
$\mathrm{0.302 \pm 0.046}$ & 
~ & ~ & ~ & 
$4.08$
\\
$\{a_2, a_4\}$ &
$\mathrm{0.279 \pm 0.047}$ & 
$\mathrm{0.189 \pm 0.060}$ & 
~& ~ & 
$0.75$
\\
$\{a_2, a_4, a_6\}$ &
$\mathrm{0.270 \pm 0.047}$ & 
$\mathrm{0.179 \pm 0.060}$ & 
$\mathrm{0.123 \pm 0.086}$ & 
~ & 
$ 0.073 $
 \\
$\{a_2, a_4, a_6, a_8\}$ &
$\mathrm{0.269 \pm 0.047}$ & 
$\mathrm{0.185 \pm 0.062}$ & 
$\mathrm{0.141 \pm 0.096}$ & 
$\mathrm{0.049 \pm 0.116}$ & 
$0.013$
\\
\hline 
\end{tabular}
\caption{The Gegenbauer moments fitted from Eq.~(\ref{eq:dispFpi3}) 
for the models~(\ref{eq:models}) of the pion twist-2 DA. 
The correlations between the moments are found at the level of 
$\approx - 15 \%$. 
}
\label{tab:an-res}
}
\end{table}

We emphasize that the fit procedure employed here
is an exploratory one, with the aim to demonstrate that the 
new method of obtaining the pion spacelike form factor works 
and being combined with LCSR provides constraints on the Gegenbauer moments. In particular, we do not take into account the 
correlations between ''experimental'' points
in Eq.~(\ref{eq:chi-sq-function}), stemming from the correlations 
within the data on the timelike form factor. 
The number of degrees of freedom (ndf) used in the fit is simply 
equal to the difference of $N_p$ and the number of Gegenbauer
moments involved in the model of the pion DA. Increasing $N_p$ 
does not influence the central values of the fitted Gegenbauer moments 
but leads to an artificial decrease of $\chi^2_{\rm min}/\mbox{ndf}$,
due to the growth of ndf. The selected $N_p$  is optimal in this respect 
and, in addition, covers the chosen interval of $Q^2$. Further
improvement of the  statistical treatment of the  relation (\ref{eq:dispFpi3})
is possible provided more detailed data on the timelike form
factor become available.

Several comments regarding the results presented in
Table 3 are in order:

\begin{itemize}
\item 
The minimal model $\{a_2\}$ with a single Gegenbauer moment
is only marginally consistent with Eq.~(\ref{eq:dispFpi3}). 
As can be seen from the right bottom plot in Fig.~\ref{fig:fit-result-a8-a6-a4-a2},
this model does not provide an accurate matching between the LCSR and 
dispersion relation at all $Q^2$. 

\item 
The simplest non-minimal model $\{a_2, a_4\}$ 
yields a satisfactory fit which becomes even better
for the  models $\{a_2, a_4,a_6\}$ and $\{a_2,a_4,a_6,a_8\}$. 

\item 
The fitted value of $a_8$ in the model $\{a_2, a_4,a_6, a_8\}$
is consistent with zero within uncertainties. 
In addition, we have probed models containing moments with $n>8$ and 
found that this tendency  persists: the fit returns small coefficients 
$a_{n\geq 8}$ with large errors, retaining practically similar 
ranges of the first three Gegenbauer moments. 
The main reason is a strong and sign alternating oscillation of 
the coefficient functions $f_n$ at large $n$. 

\item Generally, the fitted pattern of Gegenbauer moments
is in accordance with a~convergent conformal expansion (\ref{eq:confexp}).

\item 
As can be seen from Table~\ref{tab:an-res}, for all three non-minimal models 
that we consider, the fitted values of the second and fourth
Gegenbauer moment cover approximately the same intervals:
\begin{equation}
a_2 (1 \, \mbox{GeV}) = 0.22 - 0.33\,, ~~~
a_4 (1 \, \mbox{GeV}) = 0.12 - 0.25\, .
\label{eq:a2a4}
\end{equation}
We refrain from quoting central values and standard deviations instead 
of these intervals, having in  mind limitations of 
our statistical analysis.
\end{itemize}

\begin{figure}[t]\centering 
\includegraphics[scale=0.5]{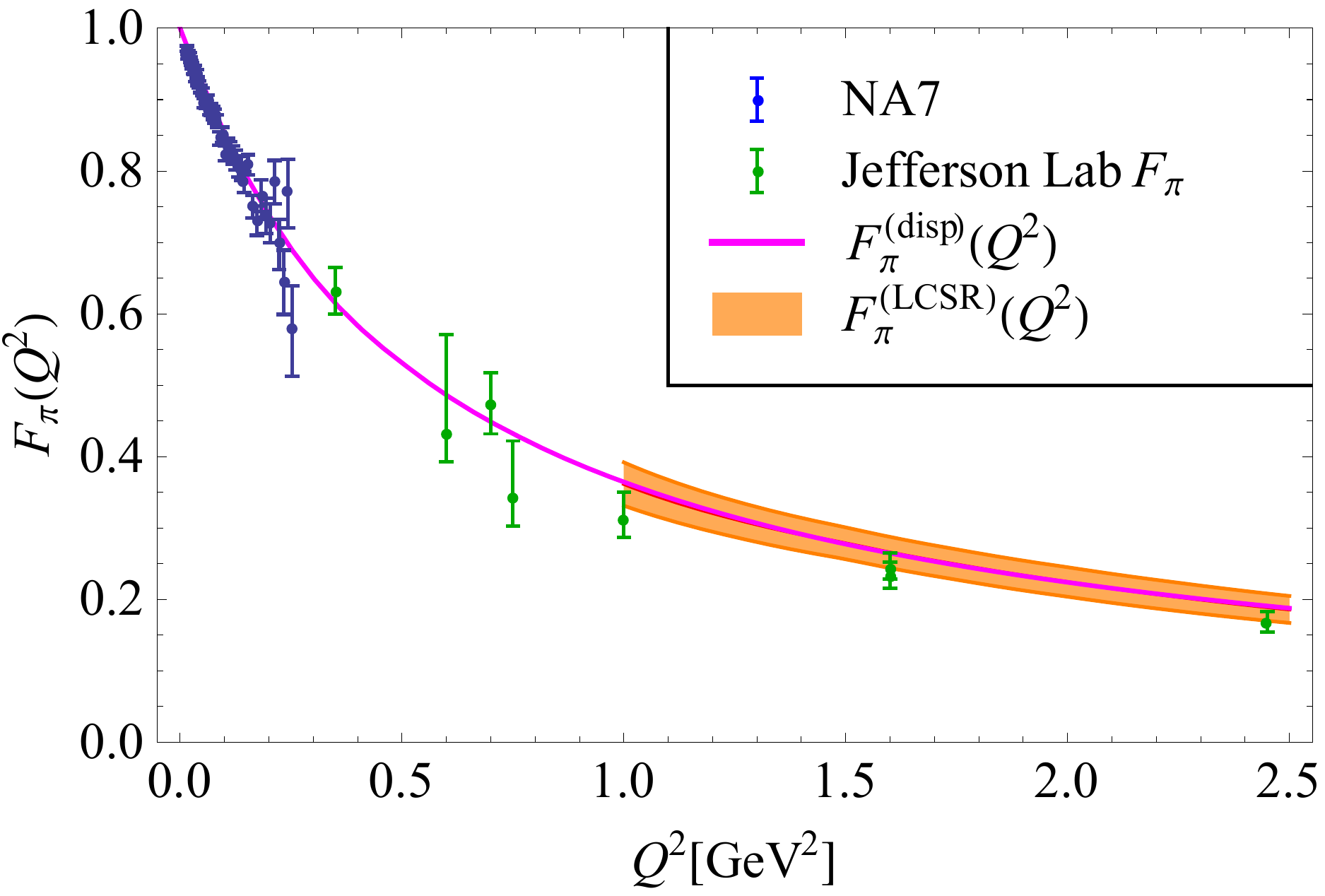}
\caption{(color online).
The pion spacelike form factor calculated from the dispersion 
relation (\ref{eq:dispFpi3}) (magenta curve; central input) 
and from the LCSR with the fitted Gegenbauer moments 
(orange band; the model $\{a_2, a_4, a_6, a_8\}$) compared with 
the measurements of NA7~\cite{Amendolia:1986wj} (blue data points)
and Jefferson Lab $F_\pi$~\cite{Huber:2008id} (green data points).}
\label{fig:lcsrvsJlab}
\end{figure}

It is instructive to compare our results with the pion spacelike form factor
measurements not used in the fit.
This comparison is displayed in Fig.~\ref{fig:lcsrvsJlab} 
where the two (correlated) results: the spacelike form factor obtained
from the dispersion relation at $Q^2=-q^2\geq 0$ and  from LCSR 
in the region (\ref{eq:region}) with the fitted Gegenbauer moments
(the model $\{a_2,a_4,a_6, a_8\}$), are plotted together with the
Jefferson Lab $F_\pi$ data 
\cite{Huber:2008id} at intermediate $Q^2$ and 
the NA7 data \cite{Amendolia:1986wj} at very small $Q^2$. 
We observe  an agreement within uncertainties and experimental errors, 
whereas at  $Q^2>1.0$ GeV$^2$ the central values of the predicted form factor 
lie slightly above the experimental points. 
\begin{figure}[t]\centering 
\includegraphics[scale=0.5]{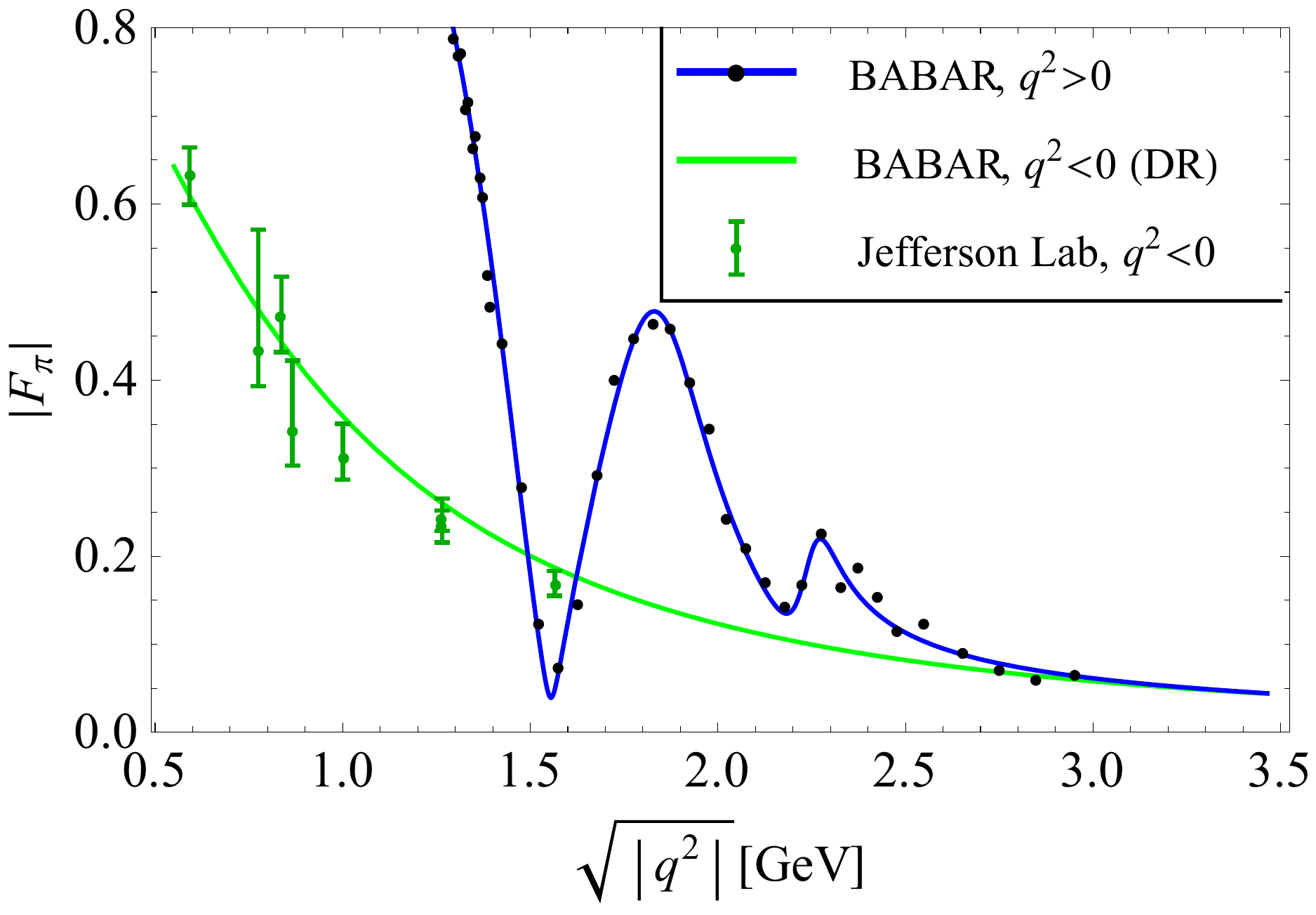}
\caption{(color online). The pion timelike form factor measured by 
the BaBar collaboration~\cite{Lees:2012cj} 
(black points (the central values) and the fit
(blue curve)) and 
the pion spacelike form factor obtained from the dispersion relation 
(\ref{eq:dispFpi2}) (green curve). 
The  Jlab data are shown by green points with error~bars. 
}
\label{fig:Fpitmlspcl}
\end{figure}

Furthermore, having at our disposal the modulus of the pion form factor
at sufficiently large timelike momentum transfer $\sqrt{s}=\sqrt{|q^2|}$, 
we compare  it with the form factor at large spacelike 
$\sqrt{Q^2}=\sqrt{|q^2|}$
inferred from Eq.~(\ref{eq:dispFpi2}).
In Fig.~\ref{fig:Fpitmlspcl} the overlap of the two form factors
at equal values of $\sqrt{|q^2|}$ is plotted.
Note that this comparison does not involve any LCSR results.
At $\sqrt{|q^2|} \gtrsim $ 3.0 GeV, 
above the region of pronounced resonances, we observe an onset of a
regime in which the timelike form factor approximately coincides 
with the spacelike one.
Generally, this equality manifests analyticity of the modulus representation. 
At the same time, since the form factor at large 
$Q^2$ is well reproduced by the LCSR based on a quark-gluon correlation
function (cf. the comparison in Fig.~\ref{fig:lcsrvsJlab}), 
it is conceivable to interpret this coincidence as the onset 
of the quark-hadron duality: calculating the form factor 
at a sufficiently large $Q^2$ in terms of the quark-gluon degrees of freedom,
we are able to predict the timelike pion form factor at $s=Q^2$.
Note on the other hand that in this region the form factor is
still far from QCD asymptotics and, as follows e.g. from LCSR, 
is dominated by soft overlap contributions. 
To see that, we use the well known asymptotic formula 
\cite{Chernyak:1977fk,Farrar:1979aw,Efremov:1979qk,Lepage:1980fj}:
\begin{equation}
\lim_{Q^2\to \infty} F_\pi(Q^2)=
\frac{8\pi\alpha_s}{9 \, Q^2}f_\pi^2
\bigg(\int\limits_0^1\!\frac{du}{u}\varphi_\pi(u,\mu)\bigg)^2,
\label{eq:Fpiasy2}
\end{equation}
(cf. Eq. (\ref{eq:Fpiasy})). 
Adopting for the pion DA, e.g., the model  $\{a_2,a_4,a_6\}$
with  Gegenbauer moments from Table~\ref{tab:an-res} and the
characteristic scale $\mu\sim \sqrt{Q^2}$, we obtain that 
in the ballpark of the ``duality'' region, 
at $Q^2 = 10 \, \mbox{GeV}^2$, the r.h.s. of the above equation 
is equal to $\simeq 0.019$,
substantially smaller than the value 
$F_\pi^{\rm (disp)} (Q^2 = 10 \, \mbox{GeV}^2)\simeq 0.05$
predicted by the dispersion relation. 

\begin{figure}[h]\centering 
\includegraphics[scale=0.4]{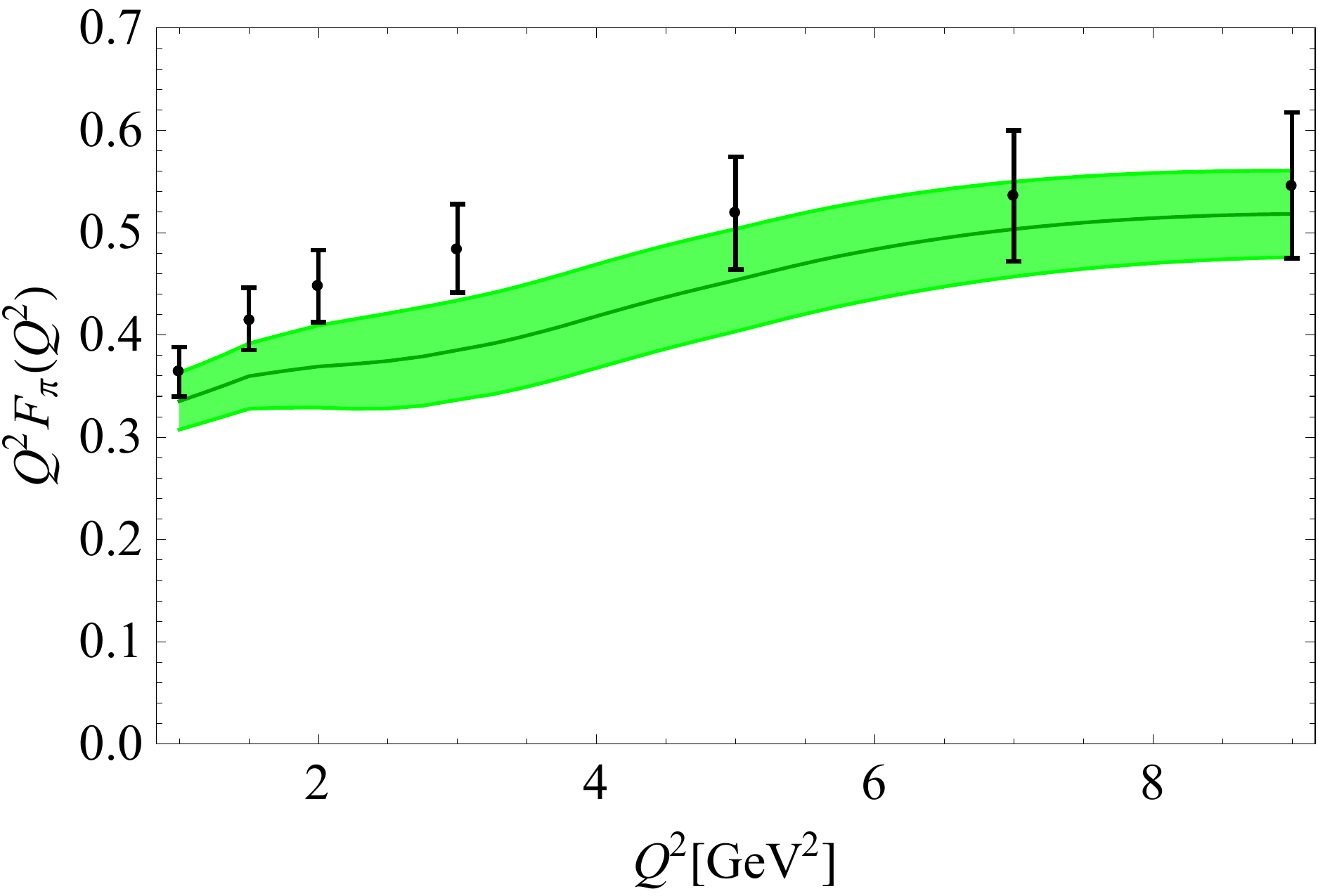}
\includegraphics[scale=0.4]{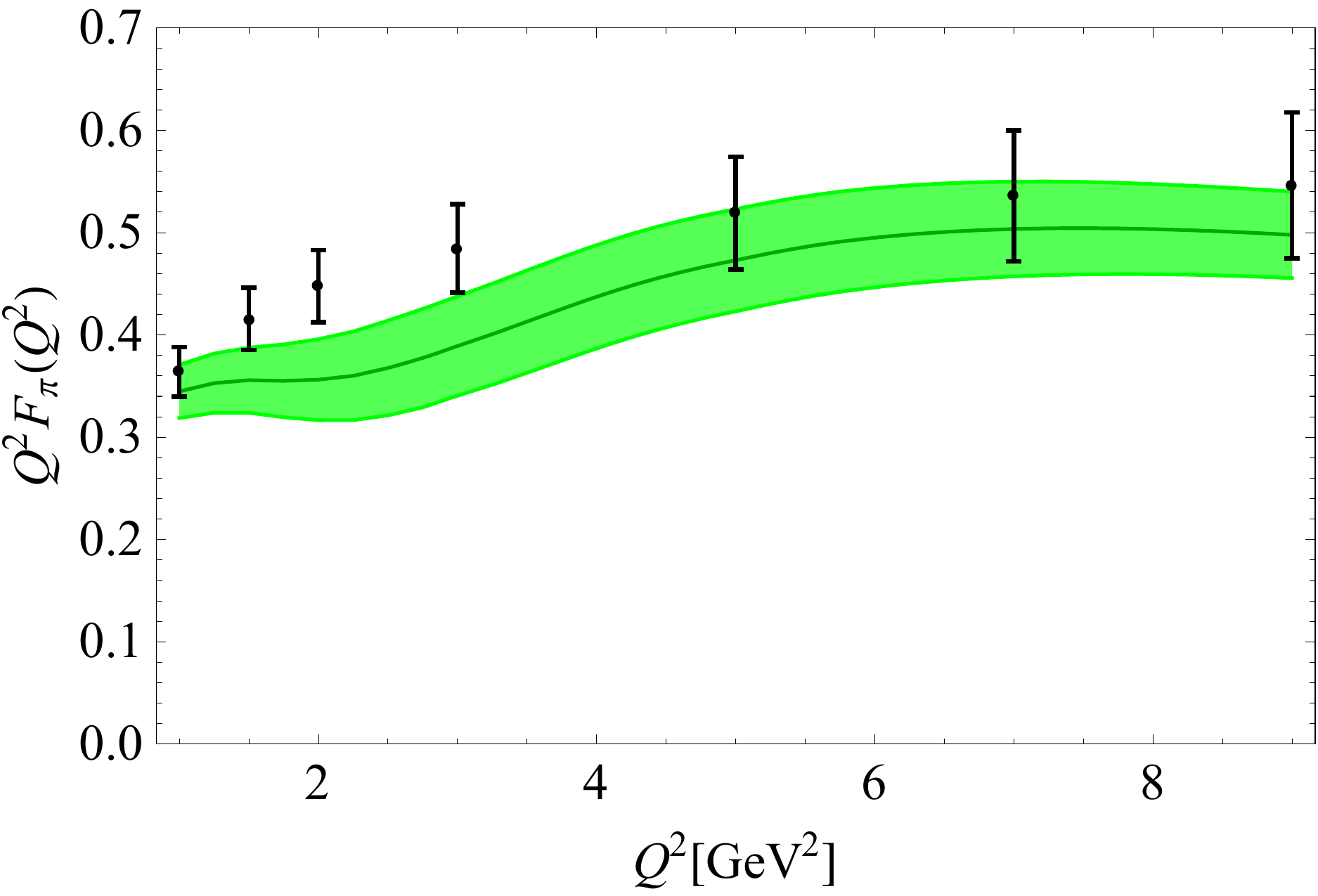}
\includegraphics[scale=0.4]{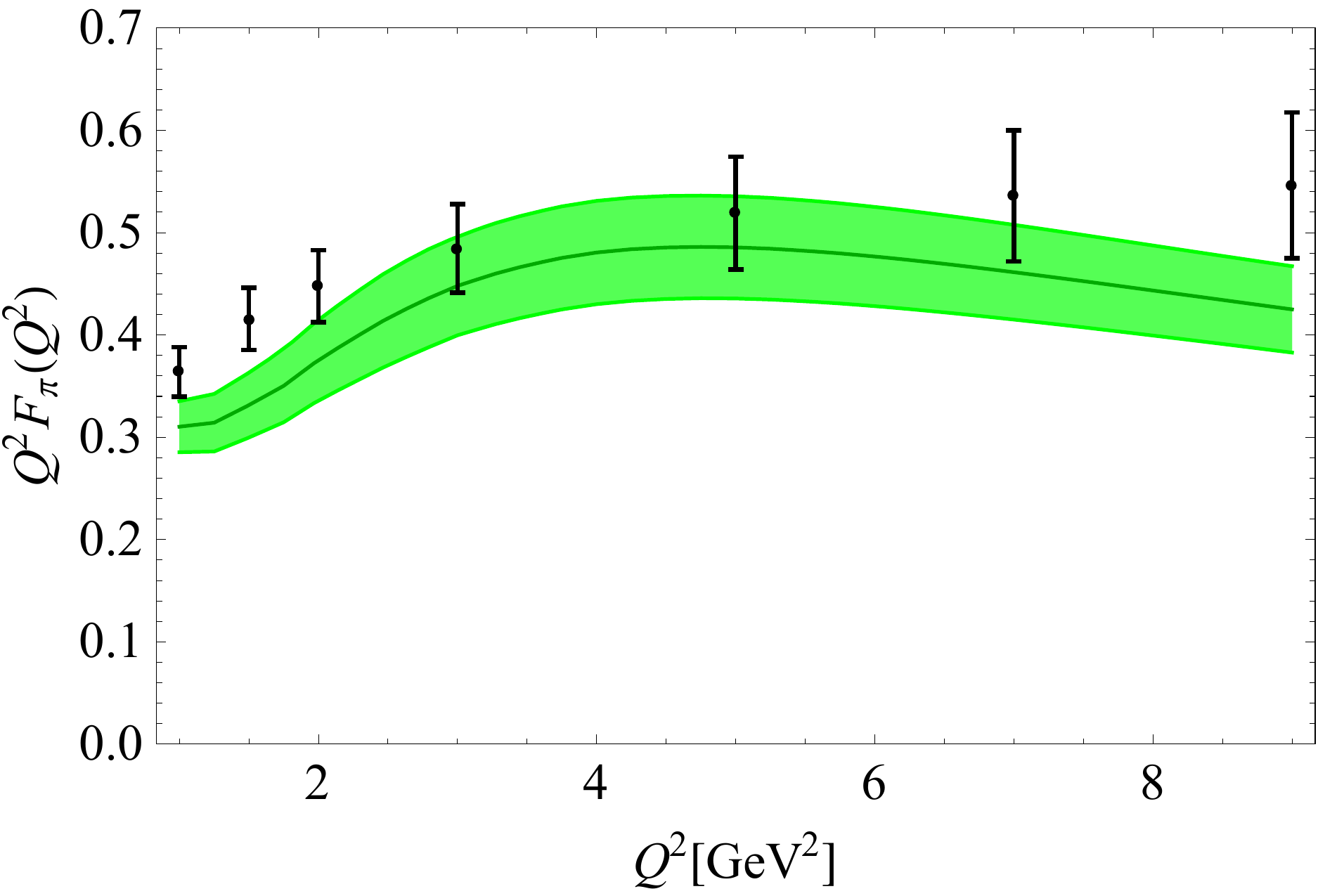}
\caption{The same as in Fig.~\ref{fig:fit-result-a8-a6-a4-a2} 
but with a fixed value of $a_2$ from Eq.~(\ref{eq:lata2}). 
The results for the models $\{a_2, a_4, a_6, a_8\}$, $\{a_2, a_4, a_6\}$, 
and $\{a_2, a_4\}$ are shown, respectively, on the top left, top right, and 
bottom  panel. 
}
\label{fig:fit-result-a6-a4-a2}
\end{figure}

Let us compare our estimates of Gegenbauer moments with the results of other methods. 
A useful compilation can be found in Ref.~\cite{Bali:2019dqc} (see Table 4 there).
We start from the determinations of $a_2$ that are independent of the experimental data 
and not correlated with higher moments. 
First of all, the moment $a_2$ is accessible in the lattice QCD.
The most accurate value has been recently obtained in Ref.~\cite{Bali:2019dqc}.
Rescaling it to the default scale:  
\begin{equation}
a_2 (1 \, {\rm GeV}) = 0.135 \pm 0.032\,,
\label{eq:lata2}
\end{equation}
we find that our fitted values in Table~\ref{tab:an-res} 
and in Eq.~(\ref{eq:a2a4}) are noticeably larger. 
Fixing $a_2$ from the lattice QCD result (\ref{eq:lata2}),   
we have repeated the fit of Eq.~(\ref{eq:dispFpi3}).
The results are given in Table~\ref{tab:an-res-B} 
(see also Fig.~\ref{fig:fit-result-a6-a4-a2}) 
for the three models $~\{a_2,a_4\}$, $\{a_2,a_4,a_6\}$, $\{a_2,a_4,a_6,a_8\}$. 
We notice that the values of $a_4$, $a_6$ increase and the quality of the fit 
becomes worse in comparison with our initial results in Table~\ref{tab:an-res}.
\begin{table}[th]
{\small 
\centering 
\begin{tabular}{|c| c |c| c|c|}
\hline 
Model &
$a_4 (1 \, {\rm GeV})$ & 
$a_6 (1 \, {\rm GeV})$ & 
$a_8 (1 \, {\rm GeV})$ & 
$\mathrm{\chi^2_{\rm min}/ndf}$ 
\\
\hline 
$\{a_2, a_4 \}$ &
$\mathrm{0.218 \pm 0.059}$ & 
~& ~ & 
$3.93$
\\
$\{a_2,a_4,a_6\}$ &
$\mathrm{0.203 \pm 0.060}$ & 
$\mathrm{0.157 \pm 0.086}$ & 
~ & 
$2.81$
 \\
$\{a_2, a_4,a_6,a_8\}$ &
$\mathrm{0.210 \pm 0.061}$ & 
$\mathrm{0.179 \pm 0.095}$ & 
$\mathrm{0.062 \pm 0.116}$ & 
$2.71$
\\
\hline 
\end{tabular}
\vspace{2mm}
\caption{The same as in Table~\ref{tab:an-res} but with a fixed 
value of $a_2$ from Eq.~(\ref{eq:lata2}).
}
\label{tab:an-res-B}}
\end{table}

Another method \cite{Chernyak:1981zz} to calculate $a_2$ 
is based on the QCD sum rules for a two-point vacuum correlation function.
Comparing our results with the estimate $a_2(1 \, \mbox{GeV})= 0.28\pm 0.08$ 
obtained by this method  in Ref.~\cite{Ball:2006wn} (see also
Ref.~\cite{Khodjamirian:2004ga}) we observe a good agreement.

The Gegenbauer moments with $n>2$ were obtained from the QCD sum rules with
nonlocal condensates, predicting 
$a_2 (1 \, {\rm GeV}) \simeq 0.20, \, a_4 (1 \, {\rm GeV}) \simeq - 0.14$ 
\cite{Bakulev:2001pa,Mikhailov:2016klg}. 
A striking difference with respect to  our estimates is the 
negative sign of the fourth Gegenbauer moment.  
The first lattice QCD values of $a_4$
obtained with the new method of Euclidean correlation function in 
Ref.~\cite{Bali:2017gfr}  still have very large uncertainties. 

A different strategy is to match an analytical expression 
for the pion e.m. form factor in terms 
of  twist-2 DA to the measured values in the spacelike
region. Certainly, the use of a hard-scattering factorization formula 
e.g.,  of the asymptotic expression in Eq.~(\ref{eq:Fpiasy2}) 
is not an adequate choice, because,  as we have convinced ourselves, 
at large but finite $Q^2$ the soft overlap part of the form factor dominates. 
Instead, in several analyses, the LCSR written in the form 
of Eq.~(\ref{eq:lcsrtot3}) was employed. 
As opposed to the method  we suggest here, this equation was directly 
fitted to the measured form factor. E.g., in  Ref.~\cite {Khodjamirian:2011ub},  
the LCSR with the model $\{a_2,a_4\}$ was fitted to the Jefferson Lab 
$F_\pi$ data \cite{Huber:2008id} 
in the region $0.6$~GeV$^2$~$<Q^2<2.45$~GeV$^2$. The results%
$$
a_2(1 \, \mbox{GeV})= 0.17\pm 0.08,~~
a_4(1 \, \mbox{GeV})= 0.06\pm 0.10
$$
have  larger uncertainties than our fit results for the same model
in Table~\ref{tab:an-res} and marginally agree with the latter within errors. 
Certainly, the use of a dispersion relation
has an advantage of giving access to a considerably larger region of 
the spacelike momentum transfer. 
\begin{table}[h]
\centering 
{\small 
\begin{tabular}{|c| c| c| c|} 
\hline 
Method &$a_2 (1 \, {\rm GeV})$ & $a_4 (1 \, {\rm GeV})$ & Ref. \\
\hline
Lattice QCD&$0.135 \pm 0.032$& -- &\cite{Bali:2019dqc}\\
QCD  sum rule &$0.28\pm 0.08$&--&\cite{Ball:2006wn}\\
QCD sum rule  with nonlocal condensate &$0.203^{+0.069}_{-0.057}$& 
$-0.143^{+0.094}_{-0.087}$&\cite{Mikhailov:2016klg,Stefanis:2020rnd}\\
LCSR fitted to Jlab data&$0.17\pm 0.08$ & $0.06\pm 0.10$ &\cite{Khodjamirian:2011ub}\\
LCSR fitted to dispersion relation
& 0.22 - 0.33
& 0.12 - 0.25
& this work\\
\hline 
\end{tabular}
}
\caption{Comparison of the second and fourth Gegenbauer moments 
obtained with various methods. }
\label{tabnew}
\end{table}

For convenience, in Table~\ref{tabnew} we put together the 
results on $a_2$ and $a_4$ obtained with various methods
and discussed above. The most important 
findings which deserve further attention and investigations are:  
1) our estimated interval for $a_2$ lies  above 
the lattice QCD prediction, and probably also 
above the result from the LCSR fitted to the Jlab data;
2) the LCSR fits prefer a positive sign of $a_4$,  
opposite to the one predicted from the sum rules with nonlocal condensates.

Gegenbauer moments are also constrained from the photon-pion transition 
form factor measured in the $\gamma^*\gamma \to \pi^0$ process.  
To calculate this form factor, one uses the method of
Ref.~\cite{Khodjamirian:1997tk}, combining  a dispersion relation
in the photon virtuality with LCSR. 
The accuracy of this method was improved in Ref.~\cite{Agaev:2010aq}
calculating important additional terms in the LCSR. 
A more recent analysis of $\gamma^*\gamma \to \pi^0$ with related methods
can be found in Ref.~\cite{Wang:2017ijn}. 
The main problem of using the photon-pion transition form factor
is a mutual discrepancy between the results of different 
experiments, especially at large $Q^2$.  
We postpone a more  detailed discussion and only mention,
for comparison, the model $\{a_2,a_4,a_6,a_8\} $ used to fit the data
in Ref.~\cite{Agaev:2010aq} (the model II): 
\begin{eqnarray}
&a_2(1 \, \mbox{GeV})=0.10-0.14,~~
&a_4(1 \, \mbox{GeV})=0.10-0.18,~~\nonumber\\
& a_6(1 \, \mbox{GeV})=0.10-0.23, ~~
&a_8(1 \, \mbox{GeV})=0.034-0.05,~~
\nonumber
\end{eqnarray}
where the upper and lower limits correspond to different experiments.
This model has the same sign pattern as our fit results 
but has a smaller second moment and does not
reveal a  convergent conformal expansion.

\section{Conclusion}
\label{sect:concl}

In this paper we suggested a new method to probe
the twist-2 pion DA by comparing two
independent ways to calculate the pion e.m. form factor in the spacelike
region. The first one employs the modified dispersion relation 
(modulus representation) in which the input is essentially provided by
the direct measurement of the pion e.m. form factor in the 
timelike region. The same form factor is calculated from the LCSR 
with a linear dependence on the Gegenbauer moments taken at a 
certain reference normalization scale.
We performed an exploratory 
numerical investigation of the equation between LCSR and
dispersion relation, employing for the latter the BaBar collaboration data 
on the pion timelike form factor. 
Adopting simple models of the pion DA, we fitted the first 
few Gegenbauer moments.  
This analysis reveals certain  gross features of the pion DA:
its form deviates from the purely asymptotic one; the minimal ansatz
with a single Gegenbauer moment $a_2$ is disfavoured and the validity of a 
converging conformal expansion is confirmed.
Our main results 
are given in Table~\ref{tab:an-res} and yield
intervals (\ref{eq:a2a4}) for the second and fourth moments.
Numerically, our prediction for $a_2$ is in the same
ballpark as the two-point QCD sum rule estimates, but exceeds
the currently most accurate lattice QCD value. 
We also found that
the spacelike form factor extracted from the dispersion relation
is consistent within errors with the Jlab data. 
Importantly, this form factor at  large momentum transfers 
$Q^2\sim $ 10 GeV$^2$ is still 
considerably larger than its perturbative QCD asymptotics.
This is in accordance with the first lattice QCD calculations
\cite{Koponen:2017fvm,Chambers:2017tuf}
of the light meson form factors at large spacelike
 momentum transfers.

Turning to future perspectives 
of the method suggested in this paper, let us first of all mention
possible modifications   
of the specific dispersion relation (modulus representation)
used here. In particular, the role of form factor zeros should be investigated.
Additional subtractions and/or differentiation in $Q^2$ can provide 
an effective suppression of the high-energy tail of the timelike form factor 
which is not directly measured. The  fit procedure applied in this 
paper can further be extended to include the data 
on the direct measurements of the spacelike form factor.  In the
timelike region, one may additionally employ the pion  
vector form factor measured in $\tau^- \to \pi^-\pi^0\nu_\tau$ decays.
Eventually, our analysis should be complemented 
with  the fit of the measured
photon-pion form factor for which the LCSR-based theoretical expression
contains the same input as the one used for the pion e.m. form
factor. 

Needless to say, the accuracy of the fit 
will  benefit from new  more accurate data on all 
types of the pion form factor. Note that 
both the region of large  $Q^2$ for the photon-pion form
factor and the region of large $s$ for the pion timelike e.m.
and vector form factors are potentially accessible at Belle II.
Finally, we foresee a perspective application of
the  same method to the  kaon DA, employing the kaon e.m. form factor 
as well as  the flavour-changing form factors  in $\tau \to K\pi\ell\nu_\ell $ decays.

\section*{Acknowledgements}
We are grateful to Irinel Caprini for useful comments.
A.K. and A.R. would like to thank the Mainz Institute for Theoretical Physics (MITP) 
of the Cluster of Excellence PRISMA+ (Project ID 39083149) for
hospitality and support. S.C. is supported by the National Science Foundation of China (NSFC) under Grant No.~11805060 and the Natural Science Foundation of Hunan Province, China (Grant No.2020JJ4160). 
The work of A.K. is supported by the Deutsche Forschungsgemeinschaft 
(German Research Foundation) under  grant 396021762 - TRR 257. 
A.R. acknowledges support by the STFC grant of the IPPP.


\section*{Appendix}
 
Here we present the expression for the Gounaris-Sakurai resonance 
formula used in Eq.~(\ref{eq:BW-model}):
\begin{equation}
{\rm BW}_{R}^{\rm GS}(s)=\frac{m_{R}^2  + m_{R} \Gamma_{R} \, d(m_{R})}
{m_{R}^2 - s + f(s, m_{R}, \Gamma_{R}) - 
i m_{R} \Gamma (s, m_{R}, \Gamma_{R})}, 
\label{eq:BW-GS}
\end{equation}
where $R = \rho, \rho', \rho'', \rho'''$  and the functions 
entering this expression are 
\begin{eqnarray}
\Gamma (s, m, \Gamma) & = & 
\Gamma \frac{s}{m^2} \left( \frac{\beta_\pi (s)}{\beta_\pi (m^2)} \right)^{\!3}, 
\qquad \beta_\pi (s) = \sqrt{1 - 4 m_\pi^2/s}, \nonumber\\
d (m) & = & \frac{3}{\pi} \frac{m_\pi^2}{k^2(m^2)} \ln \left(\frac{m+2 k (m^2)}{2 m_\pi} \right) 
+ \frac{m}{2 \pi k (m^2)} - \frac{m_\pi^2 m}{\pi k^3 (m^2)}, \nonumber\\
f (s, m, \Gamma) & = & \frac{\Gamma m^2}{k^3 (m^2)} 
\left(k^2 (s) (h(s) -h (m^2) + (m^2 -s) k^2 (m^2) h^\prime (m^2)  \right), \nonumber\\
k (s) & = & \frac{1}{2} \sqrt{s} \beta_\pi (s), \nonumber\\
h (s) & = & \frac{2}{\pi} \frac{k(s)}{\sqrt{s}} 
\ln \left(\frac{\sqrt{s} + 2 k (s)}{2 m_\pi} \right), \qquad h^\prime 
            (s) = \frac{d h(s)}{d s}. 
\nonumber 
\end{eqnarray}
and the KS representation used for the $\omega$-resonance 
is given by 
\begin{equation}
{\rm BW}_\omega^{\rm KS} (s, m_\omega, \Gamma_\omega) = 
\frac{m_\omega^2}{m^2_\omega - s - i \, m_\omega \Gamma_\omega} \,.
\end{equation}

\end{document}